\documentclass[%
reprint,
superscriptaddress,
nofootinbib,
twocolumns,
notitlepage,
amsmath,amssymb,
aps,
prx
]{revtex4-2}

\usepackage{graphicx}% Include figure files
\usepackage{dcolumn}% Align table columns on decimal point
\usepackage{bm}% bold math
\usepackage{tablefootnote}

\usepackage{amsthm}
\usepackage{mathtools}
\usepackage{amsmath}
\usepackage{amssymb}
\usepackage{graphicx}
\usepackage{url}
\usepackage{float}
\usepackage{bm}
\usepackage{ifthen}
\usepackage[usenames,dvipsnames]{color}
\usepackage{mathrsfs}
\usepackage[colorlinks=true,citecolor=blue,urlcolor=black]{hyperref}
\usepackage{float}
\usepackage{subfigure}
\usepackage{enumitem}
\usepackage{color}
\usepackage{setspace}
\usepackage{braket}
\usepackage{comment}
\usepackage{lipsum}
\newcommand{\be}{\begin{equation}}
\newcommand{\ee}{\end{equation}}

\begin{document}

\title{Fully Passive Twin-Field Quantum Key Distribution}

\author{Wenyuan Wang}
\thanks{wenyuan.wang@ucalgary.ca. Present address: Department of Physics, University of Calgary, Alberta, T2N 1N4, Canada.}
\affiliation{Department of Physics, University of Hong Kong, Pokfulam Road, Hong Kong}

\author{Rong Wang}
\affiliation{Department of Physics, University of Hong Kong, Pokfulam Road, Hong Kong}

\author{H. F. Chau}
\affiliation{Department of Physics, University of Hong Kong, Pokfulam Road, Hong Kong}

\author{Hoi-Kwong Lo}
\thanks{hklo@comm.utoronto.ca}
\affiliation{Department of Physics, University of Hong Kong, Pokfulam Road, Hong Kong}
\affiliation{Dept. of Physics, University of Toronto, Toronto,  Ontario, M5S 3G4, Canada}
\affiliation{Dept. of Electrical \& Computer Engineering, University of Toronto, Toronto,  Ontario, M5S 1A7, Canada}
\affiliation{Centre for Quantum Information and Quantum Control (CQIQC), Toronto,  Ontario, M5S 1A7, Canada}
\affiliation{Quantum Bridge Technologies, Inc., 100 College Street, Toronto, ON M5G 1L5, Canada}

\begin{abstract}
	
	We propose a fully passive twin-field quantum key distribution (QKD) setup where basis choice, decoy-state preparation and encoding are all implemented entirely by post-processing without any active modulation. Our protocol can remove the potential side-channels from both source modulators and detectors, and additionally retain the high key rate advantage offered by twin-field QKD, thus offering great implementation security and good performance. Importantly, we also propose a post-processing strategy that uses mismatched phase slices and minimizes the effect of sifting. We show with numerical simulation that the new protocol can still beat the repeaterless bound and provide satisfactory key rate.
	
\end{abstract}

\date{\today}
\maketitle

\textbf{Background.} For a practical quantum key distribution (QKD) system, active modulation devices in a source may introduce side-channels \cite{IMleak1,IMleak2} or be susceptible to Trojan horse attacks \cite{trojan,trojan2}, leaking information to Eve. Therefore, it would be ideal if we can remove the modulation devices in the source and implement decoy-state setting choice \cite{decoystate_Hwang,decoystate_LMC,decoystate_Wang} and the encoding of signals passively by post-selection or post-processing only.

Passive decoy-state \cite{passivedecoy1,passivedecoy2} and passive encoding \cite{passiveEncoding} schemes have both been proposed for BB84 systems. More recently, fully passive BB84 schemes \cite{FullyPassiveThis,FullyPassiveAlternative} combining both passive decoy-state preparation and encoding have also been proposed, which entirely remove active modulators in the source. Two recent experimental demonstrations of fully passive BB84 have been reported successfully \cite{PassiveExperiment01,PassiveExperiment02}. Passive state-preparation has also been implemented for Gaussian-modulated continuous-variable (CV) QKD \cite{passiveCV} and very recently for discrete-modulated CV-QKD \cite{passiveDMCV}.

All the above passive schemes, however, aim at removing side-channels in source modulators. We can further improve the implementation security of the system if we can apply passive encoding and decoy-state choice to measurement-device-independent (MDI) protocols \cite{mdiqkd,TFQKD}, which can remove side-channels in detectors. Particularly, the new twin-field (TF) QKD \cite{TFQKD} protocol can provide both measurement-device-independence and exceptionally high key rate. In fact it can beat the upper bound for the rate-distance trade-off relation without a repeater \cite{PLOB} and has enabled QKD over as far as 1000 kilometers \cite{TFexperiment08}. Ever since its proposal, TF-QKD has led to much theoretical interest \cite{TFQKD01,TFQKD02,TFQKD03,TFQKD04,simpleTFQKD,TFQKD05,TFQKD06,asymTFQKD,asymTFQKD2} in recent years and many experimental demonstrations \cite{TFexperiment01,TFexperiment02,TFexperiment03,TFexperiment04,injectionlocking1,TFexperiment05,injectionlocking2,TFexperiment06,TFexperiment07,asymTFQKDexperiment}. While there has been a recent proposal for passive decoy-state TF-QKD \cite{passiveTF}, it still requires active modulators for encoding. {\color{black}Importantly, the fully passive BB84 schemes \cite{FullyPassiveThis,FullyPassiveAlternative} cannot be applied here, as they are based on two-mode (e.g. polarization) encoding, while TF-QKD uses single-mode phase encoding}. So far, there has never been any passive-encoding (or fully passive) scheme for TF-QKD.

\begin{figure}[t]
	\includegraphics[scale=0.34]{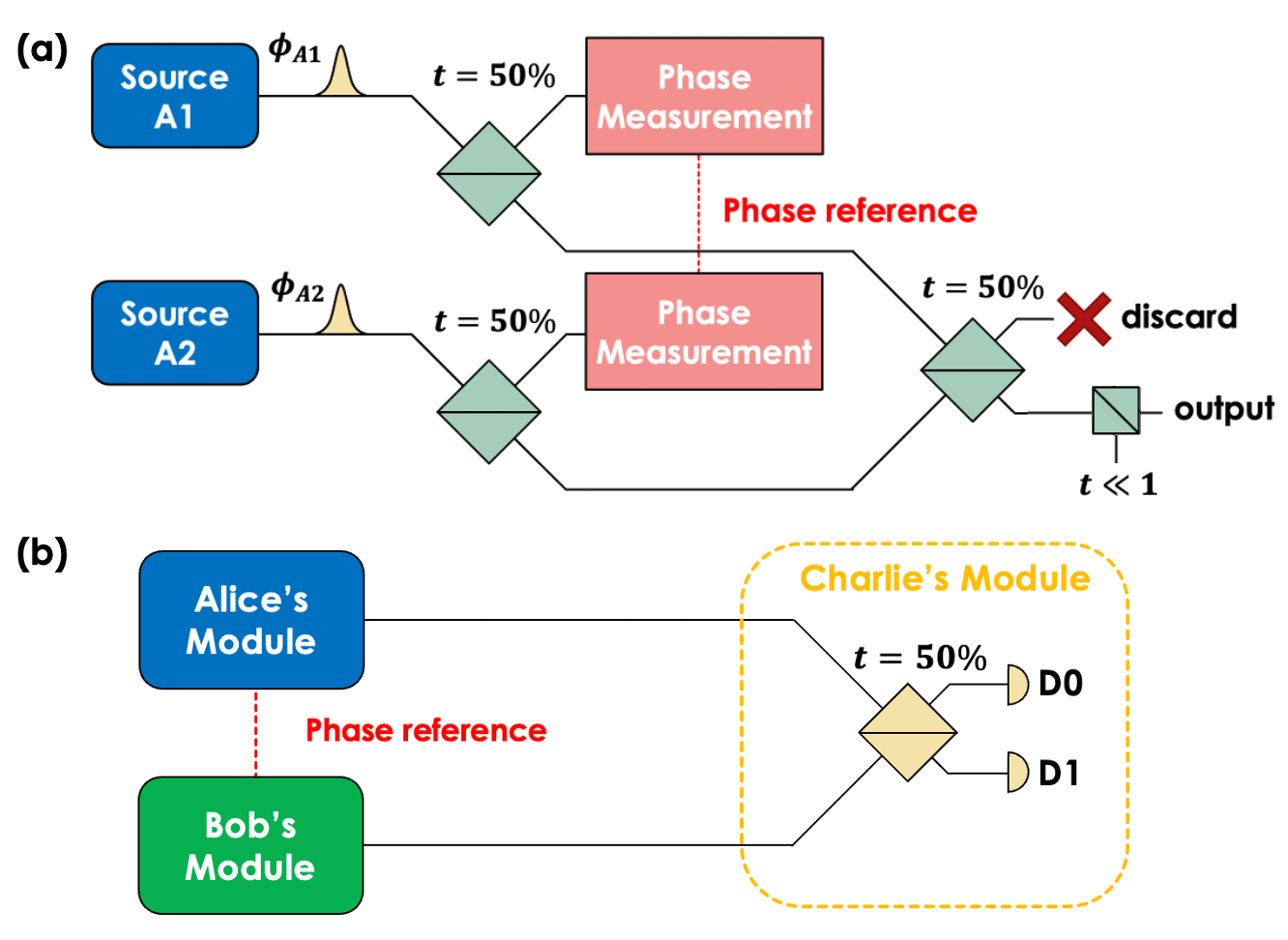}
	\caption{(a) Alice's fully passive source module (Bob holds an identical module with sources B1, B2). She prepares two strong laser pulses and splits off each of them respectively to measure the phases $\phi_{A1}$ and $\phi_{A2}$. She then lets the two pulses interfere and attenuates one of the output ports to single-photon level before sending off the signal to Charlie. Alice can optionally replace the ``discard" port with a classical photodiode, but the detection results are already implied by the phases. (b) The full setup where Alice and Bob prepare signals and send them to Charlie, who performs a swap test and publicly announces the results. Alice and Bob also need a common phase reference, e.g. established using strong pulses.}
	\label{fig:1}
\end{figure} 

In this work, we build upon the CAL TF-QKD protocol \cite{simpleTFQKD} and propose the first fully passive TF-QKD scheme, where all the encoding, decoy-state setting choice, as well as even the basis choice, are entirely generated by post-processing, and all the degrees-of-freedom (DOFs) come from the phase randomness of the laser sources themselves. Importantly, our protocol removes source modulator and detector side-channels at the same time. Additionally, to address the main challenge of heavy sifting (which would occur if one performs naive post-selection), we also propose a post-processing strategy that fully utilizes mismatched phase slices, which greatly improves sifting efficiency. This allows us to have a protocol with very high implementation security while maintaining satisfactory key rate.

\textbf{Setup.} In the passive setup, each of Alice and Bob holds two independent laser sources A1 and A2 (B1 and B2). The pulses are first generated at strong intensity levels, allowing Alice and Bob to split off the pulses and measure classically the phases of all pulses (e.g. by interfering with a reference laser), denoted as $\phi_{A1},\phi_{A2},\phi_{B1}$ and $\phi_{B2}$. For simplicity, we set the four laser sources to all generate pulses at the same intensity $\mu_{max}/2$. 

The most crucial assumption here is that the phases are all \textit{independent} and \textit{random}, i.e. uniformly distributed between $[0,2\pi)$. Such inherent phase randomness in laser sources has already been used for QKD systems and quantum random number generators (QRNGs) \cite{phaserandom1,QRNG1,QRNG2,QRNG3,QRNG4}. We also assume that Alice and Bob can establish a common phase reference for all their measurements.

Alice (Bob) first lets A1 and A2 (B1 and B2) interfere. The signal from one output port is then attenuated to single-photon level and sent to Charlie, who lets the two incoming signals interfere and announces the click events from detectors D0 and D1. Alice (Bob) can optionally observe the intensity at the other port, but this information is already contained in $|\phi_{A1}-\phi_{A2}|$ ($|\phi_{B1}-\phi_{B2}|$). The setup can be seen in Fig. 1.\\

\begin{figure}[h]
	\includegraphics[scale=0.201]{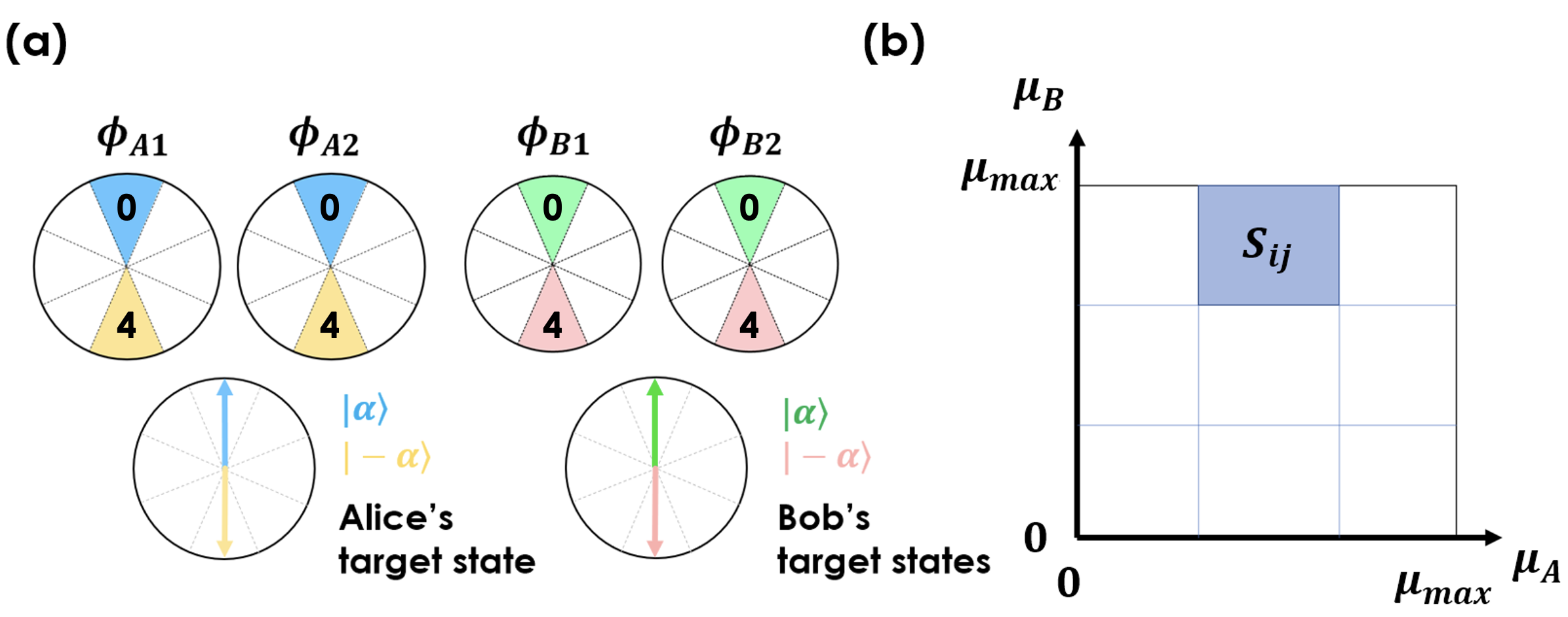}
	\caption{(a) The post-selection of phase slices in the key-generating X basis. Alice and Bob each only post-select cases where their two local slices are the same and have indices of $k=0$ or $k=N/2$ (corresponding to $0$ and $\pi$ phases). Here the total slice number is $N=8$. They can additionally use mismatched slices to generate key, which we will later explain. (b) The post-selection of decoy intensities in the Z basis. Alice and Bob can e.g. divide the domain into a $3\times 3$ grid where each $S_{ij}$ corresponds to a decoy setting.}
	\label{fig:2}
\end{figure} 

\textbf{Protocol.} We can use the above setup to implement fully passive CAL TFQKD \cite{simpleTFQKD}:\\

1. Alice and Bob prepare the states based on the setup in Fig. \ref{fig:1} and send out the signals without any modulation, while recording the locally measured $\phi_{A1},\phi_{A2},\phi_{B1}$ and $\phi_{B2}$. Charlie announces the detection results $k_c, k_d$ from detectors D0 and D1.

2. {\color{black}During post-processing, Alice randomly chooses between a coding phase (X basis) and a decoy phase (Z) based on local random classical bits. Then, she announces the random choices to Bob. Note that, since basis choice happens \textit{after} the transmission, there is no need for basis sifting in fully passive TF-QKD.}

(1) In the signal X basis, Alice and Bob divide up the $[0,2\pi)$ domain of phases into N slices (N being an even number) and assign each of the phases they measured into a slice indexed by $k$ (we denote the indices by $k_{A1},k_{A2},k_{B1}$ and $k_{B2}$). In the simplest form of post-selection, Alice (Bob) only announces a successful event if the two local slices $k_{A1},k_{A2}$ ($k_{B1},k_{B2}$) both happen to be the same \textit{and} have an index of $k=0$ or $k=N/2$, as shown in Fig. \ref{fig:2} (a). The event is discarded if either Alice or Bob fails. They each keep their slice indices secret (corresponding to their classical bits $0$ or $1$).

(2) In the decoy Z basis, Alice and Bob each post-select their signals based on $|\phi_{A1}-\phi_{A2}|$ and $|\phi_{B1}-\phi_{B2}|$, which is equivalent to measuring and post-selecting the output intensities $\mu_A$ and $\mu_B$, both of which randomly lie between $[0,\mu_{max}]$ following an intensity probability distribution $p(\mu)=1/(\pi \sqrt{\mu(\mu_{max}-\mu)})$ \cite{FullyPassiveThis}. Alice and Bob can simply divide the ranges into continuous post-selection regions $S_{ij}$ (similar to the passive decoy-state method \cite{passivedecoy1}) and announce the region each signal falls in. For instance, they can divide the domain equally into the $3\times 3$ square regions shown in Fig. \ref{fig:2} (b), which conceptually correspond to the intensity set of $\{\mu,\nu,\omega\} \times \{\mu,\nu,\omega\}$ for active TF-QKD, and perform decoy-state analysis to estimate photon-number yields $Y_{mn}$.

3. Alice and Bob perform error-correction on their raw key and privacy amplification, the amount of which is estimated using the upper-bounded phase error rate and the classical information leakage, same as for active TF-QKD.\\

While the above protocol is already a functioning fully passive CAL TF-QKD scheme, its main drawback would be the heavy sifting in the X basis (there is only a low probability of $4/N^4$ for all slices to match the pattern in Fig. \ref{fig:2} (a)). To address this problem, we propose a post-processing strategy where Alice and Bob match up the originally-discarded slices to improve sifting. {\color{black}Any combination $\{k_{A1},k_{A2},k_{B1},k_{B2}\}$ can be potentially used to generate key, not just the $0$ or $\pi$ positions. The mismatches of slice positions might correspond to higher QBER, lower output intensity, or misalignment between Alice and Bob, but these can all be treated as channel noise and do not affect the security of the protocol. Of course, some highly mismatched combinations might have zero key rate, but they will not reduce the total key rate since we can choose only the combinations with positive key rate to generate key.
	
	More specifically, here Alice and Bob match up the slices into opposite-facing slice pairs, which we index by $\{l_{A1},l_{A2},l_{B1},l_{B2}\}$. Each index $l_i$ represents a pair of slices at position $k_i=l_i$ or its opposite $k_i=(l_i+N/2) \bmod N$ that differs by a phase $\pi$. Alice and Bob each randomly post-select from two opposite patterns, e.g. for Alice, she post-selects $\{k_{A1}=l_{A1},k_{A2}=l_{A2}\}$ and the opposite pair $\{k_{A1}=(l_{A1}+N/2) \bmod N, \: k_{A2}=(l_{A2}+N/2) \bmod N\}$ to represent bits $0$ or $1$. Therefore, each set of indices $\{l_{A1},l_{A2},l_{B1},l_{B2}\}$ represents a ``basis" to generate key, and the basis position defining two possible slice pairs (but not the actual pair selected) is publicly announced to Eve. The final key rate is simply a summation of the key rate from all basis combinations:
	
	\begin{equation}\label{eq:rate}
		\begin{aligned}
			R =& {4 \over {N^4}} \sum_{l_{A1}=0}^{(N/2)-1} \sum_{l_{A2}=0}^{N-1} \sum_{l_{B1}=0}^{(N/2)-1} \sum_{l_{B2}=0}^{N-1}\\
			&[\max(0,R_{0,1} (l_{A1},l_{A2},l_{B1},l_{B2})) \\
			&+ \max(0,R_{1,0} (l_{A1},l_{A2},l_{B1},l_{B2}))],
		\end{aligned}
	\end{equation}
	
	\begin{figure}[t]
		\includegraphics[scale=0.19]{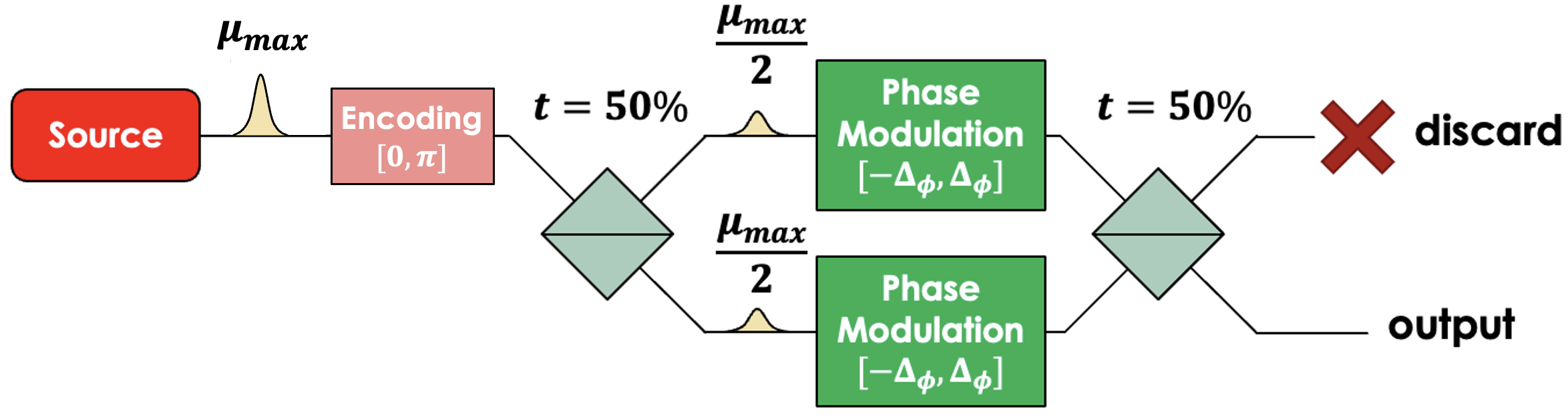}
		\caption{An equivalent local quantum channel $\mathcal{E}_A$ inside Alice's lab. In the X basis, it perturbs perfectly encoded signal states into the physical output signal from the setup in Fig. \ref{fig:1} (a). Same applies to Bob's lab.}
		\label{fig:3}
	\end{figure} 
	
	\begin{figure}[t]
		\includegraphics[scale=0.335]{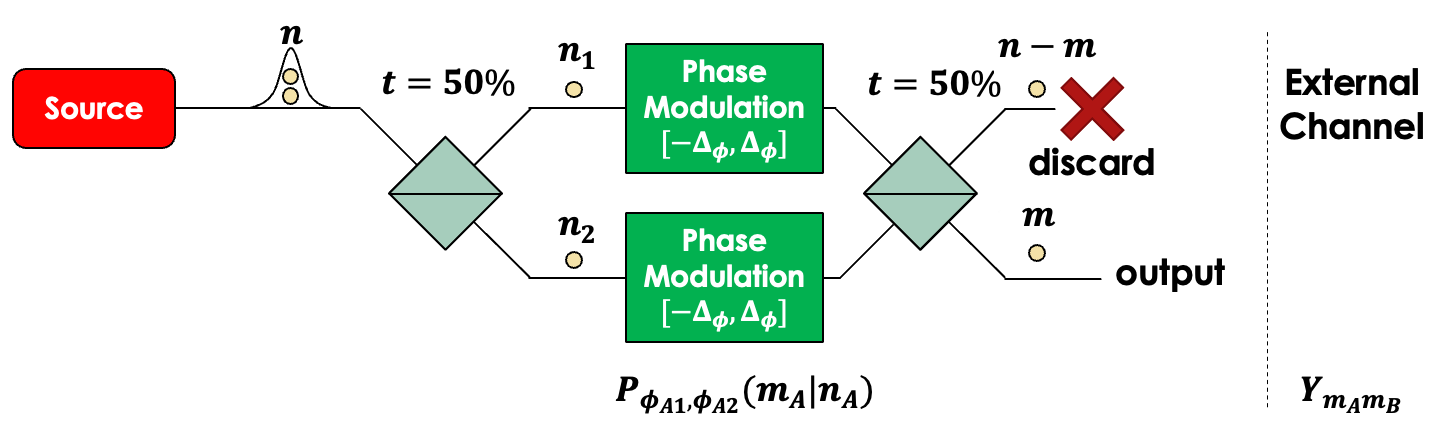}
		\caption{{\color{black}In the Z basis, for a Fock state, the local channel $\mathcal{E}_A$ is equivalent to a local loss that reduces $n_A$ photons to $m_A$ with probability of $P_{\phi_{A1},\phi_{A2}}(m_A|n_A)$. Same applies to Bob. From decoy-state analysis, Alice and Bob obtain yields $Y_{m_Am_B}$ for only the external channel. They multiply them by the transmittances of local channels to obtain the would-be yields through both the local and external channels, $Y'_{n_An_B}$. The Cauchy-Schwarz inequality is then applied at the source.}}
		\label{fig:3b}
	\end{figure} 
	\noindent which includes a sifting factor of $4/N^4$ corresponding to the probability of signals falling within each pattern of pairs of slices. Note that here $l_{A1}$ and $l_{B1}$ range from $0$ to $N/2-1$, but $l_{A2}$ and $l_{B2}$ range from $0$ to $N-1$,  to avoid duplication of phase pairs \cite{footnote2}. The subscripts of $R_{k_c,k_d}$ correspond to Charlie's detection patterns. It is possible to further reduce this factor to $2/N^3$ using rotational symmetry of global phase. More details on the sifting strategy can be found in Supplemental Material, Sec. D.}

Importantly, since we add up all combinations of phase slices, increasing the slice number $N$ does not affect the sifting factor and only increases the key rate. Our slice size is only limited by the accuracy and resolution of the classical local detection (at least in the asymptotic scenario with infinite data size, which is the focus of this work). This is the main reason why, as we will later show, our key rate is only moderately lower than the active counterpart, although we are slicing and matching the phases from four independent sources.

\textbf{Security.} The main difference between the fully passive scheme and its active counterpart is the finite size of the phase slices, $\Delta_\phi=\pi/N$ (for convenience, we define a slice to be $2\Delta_\phi$). This is a source preparation imperfection, which results in (1) the signal state QBER being inherently higher due to the slightly mismatched phases of Alice and Bob's signals, even if they choose the same slice, and (2) the characterization of the source states (which are just $\ket{\alpha},\ket{-\alpha}$ for active CAL TF-QKD) being imperfect, since the interference between sources (e.g. A1 and A2) with finite phase slices results in fluctuations in both the intensity and the encoding (phase). The latter point may result in a security loophole and needs more careful treatment.

{\color{black}Here we consider a local quantum channel \cite{channel1,channel2,channel3} $\mathcal{E}_A$ ($\mathcal{E}_B$) inside Alice's (Bob's) lab, as shown in Fig. \ref{fig:3}. The channel accepts a signal, splits it in two and applies a random phase modulation between $[-\Delta_\phi,\Delta_\phi]$ on each path. It then recombines and interferes the signals from the two paths. In the X basis, perfectly encoded $\ket{\alpha},\ket{-\alpha}$ states going through this quantum channel will result in the exact same output statistics as that of the physical source in Fig. \ref{fig:1} (a) after Alice post-selects slice pairs $l_{A1}=0$ and $l_{A2}=0$ \cite{footnote3}.
	
	We consider an equivalent virtual protocol where Alice (Bob) prepares perfectly encoded states ($\ket{\alpha},\ket{-\alpha}$ in the X basis, cat states in the Z basis) and sends them through first the local channel $\mathcal{E}_A$ ($\mathcal{E}_B$) and then the external channel. Just like for active CAL TF-QKD, they can bound the cat state statistics by sending Fock states through the same three channels and applying the Cauchy-Schwarz inequality.
	
	For the passive scenario, we can pessimistically ``attribute" the local quantum channels to Eve and consider them as part of the environment, only if the local channels are basis-independent. The problem is, for the Z basis, we estimate the yields $Y_{m_Am_B}$ by performing decoy-state analysis on observed WCP state statistics, but they only represent the effect of the external physical channel and do not include $\mathcal{E}_A$ and $\mathcal{E}_B$. Importantly, we observe that, for Fock states, the local channels can be represented as simply local losses, as shown in Fig. \ref{fig:3b}, which reduce e.g. $n_A$ photons to $m_A$ photons with probability $P_{\phi_{A1},\phi_{A2}}(m_A|n_A)$. We write the ``corrected yields" as:
	
	\begin{equation}\label{eq:yield}
		\begin{aligned}
			&Y'_{n_An_B} = \int_{-\Delta_\phi}^{\Delta_\phi} \int_{-\Delta_\phi}^{\Delta_\phi} \int_{-\Delta_\phi}^{\Delta_\phi} \int_{-\Delta_\phi}^{\Delta_\phi} \\
			& \sum_{m_A=0}^{n_A}\sum_{m_B=0}^{n_B}P_{\phi_{A1},\phi_{A2}}(m_A|n_A)P_{\phi_{B1},\phi_{B2}}(m_B|n_B) Y_{m_Am_B} \\
			&d\phi_{A1}d\phi_{A2}d\phi_{B1}d\phi_{B2},
		\end{aligned}
	\end{equation}
	
	\noindent which represent the would-be statistics of Fock states in the Z basis sent from a perfect source and going through $\mathcal{E}_A$, $\mathcal{E}_B$ and the external channel (same as signals in the X basis, thus preserving basis-independence of $\mathcal{E}_A$, $\mathcal{E}_B$). We can then bound the phase error rate in the same fashion as an active TF-QKD protocol with perfect sources.} The detailed derivation of corrected yields can be found in Supplemental Material, Sec. C.

\begin{figure}[h]
	\includegraphics[scale=0.225]{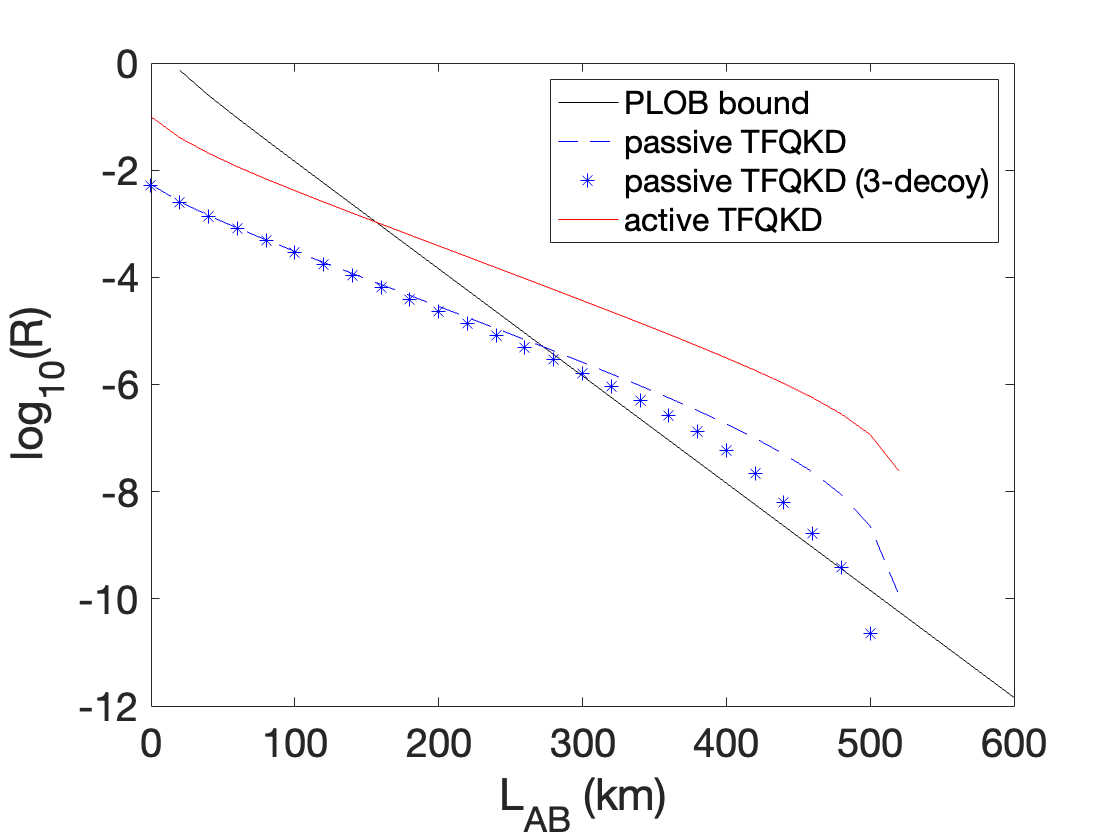}
	\caption{Simulation of the key rate for passive TF-QKD (infinite decoys, or three decoys for each of Alice and Bob) versus active TF-QKD. Here we only consider the scenario of infinite data size. The PLOB bound \cite{PLOB} is included for comparison. As can be seen, passive TF-QKD only has moderately lower key rate than the active counterpart and can still surpass the PLOB bound under practical settings.}
	\label{fig:4}
\end{figure} 

\textbf{Results.} Here we perform a simple simulation for passive TF-QKD versus its active counterpart in the asymptotic limit of infinite data size and infinite decoys (i.e. we assume perfectly estimated $Y_{m_Am_B}$). We set a dark count rate of $p_d=10^{-8}$, no misalignment \cite{footnote4}, fiber loss of $0.2dB/km$, detector efficiency of $\eta_d=1$ (the actual efficiency can be incorporated into channel loss) and an error correction efficiency of $f_e=1$. The signal intensity is optimized (approximately $\mu_{max}=0.016$ to $0.03$ depending on distance). We set the slice number to a reasonably large $N=24$. More discussions on choosing N can be found in Supplemental Material Sec. G.

We also plot the key rate with three passive decoy settings for each of Alice and Bob (using the $3\times3$ grid shown in Fig. \ref{fig:2} (b)). The signal intensity and the grid partition are both optimized. More details on the decoy analysis models used in the simulation can be found in Supplemental Material, Sec. E.

As can be seen in Fig. \ref{fig:4}, fully passive TF-QKD yields satisfactory key rate and exceeds the repeaterless bound \cite{PLOB}. The asymptotic key rate is about 1.2 orders-of-magnitude lower than the active counterpart, due to the inherent QBER and increased local losses resulting from finite phase slices, in exchange for the much better implementation security. The key rate using three decoy settings is lower than that of the infinite-decoy case at longer distances, mainly due to decoy intensities being relatively small (since all decoy intensities are no larger than $\mu_{max}$), although it can still beat the PLOB bound.

{\color{black}
	Additionally, we include a preliminary finite-size analysis in Supplemental Material Sec. I to show that fully passive TF-QKD offers decent key rate even in the finite-key regime, demonstrating the practicality of our proposal.
}

Nonetheless, note that the main purpose of our scheme is not to reach a record-breaking key rate, but rather to provide a higher implementation security than either active BB84 or active TF-QKD (while still maintaining good key rate). It would be useful not only at long distances, but even at close-to-medium distances (where it does not beat the PLOB bound), such as in a metropolitan network with untrusted relays and multiple users.

\textbf{Discussions.} In this work, we have proposed a fully passive TF-QKD protocol that removes both detector and source modulator side-channels while also offering satisfactory key rate.

Our fully passive scheme is also potentially extendable to other variants of TF-QKD, e.g. no-phase-post-selection (NPP) TF-QKD \cite{TFQKD05} (which physically uses the same signals) or phase-matching (PM) QKD \cite{TFQKD02}, which may be the subjects of future work.

A main challenge for implementing the scheme experimentally would be maintaining phase/frequency stability across all four lasers. Stability between local pairs of sources can be addressed by using a time-delay scheme \cite{FullyPassiveThis} that allows Alice and Bob to each just use a single laser. The bigger challenge, however, is to remotely establish the global phase reference and maintain frequency stability between Alice and Bob. This is also a challenge for active TF-QKD schemes, which usually use phase-locking schemes that send reference signals through service fibers to Alice and Bob respectively, implementing optical phase locked loops \cite{TFexperiment03,TFexperiment04,TFexperiment05,TFexperiment06,TFexperiment07}, time-frequency metrology \cite{TFexperiment02,TFexperiment07}, or injection locking \cite{injectionlocking1,injectionlocking2} (categorization from Ref. \cite{nophaselocking1}). However, a unique challenge for passive TF-QKD is that it needs independent phase randomness between Alice's and Bob's sources, making phase-locking no longer viable. In principle, though, Alice and Bob can send strong signals to Charlie to evaluate their phase drift and use classical feedback or post-processing mechanisms to maintain the frequency stability \cite{footnote6}. There have very recently been proposals for TF-QKD without phase-locking \cite{nophaselocking1,nophaselocking2}. Inspired by \cite{nophaselocking2}, we propose one possible frequency drift compensation scheme in Supplemental Material, Sec. H.\\

\textbf{Acknowledgments.} We thank Guillermo Currás-Lorenzo, Marcos Curty, Victor Zapatero, Chengqiu Hu, Li Qian, Chenyang Li, Zhen-Qiang Yin and Shuang Wang for helpful discussions. The authors thank financial support by NSERC, MITACS, CFI, ORF, Huawei Technologies Canada, Inc., the Royal Bank of Canada, the University of Hong Kong start-up grant. Wenyuan Wang thanks the Hong Kong RGC General Research Fund and the University of Hong Kong Seed Fund for Basic Research for New Staff.

\appendix

\tableofcontents

\section{Comparison with Fully Passive BB84}

Previously, we have proposed a fully passive BB84 scheme \cite{FullyPassiveThis}, which can impressively remove active modulators for both decoy-state choice and encoding. \footnote{Another work \cite{FullyPassiveAlternative} that uses an alternative decoy-state analysis and security proof has also been reported. Also, there have been two successful experimental demonstrations \cite{PassiveExperiment01,PassiveExperiment02} of fully passive BB84.}

However, Ref. \cite{FullyPassiveThis} focuses on a two-mode encoding setup (which can be either polarization-encoding or equivalently time-bin phase encoding), and it post-selects from states spanning the entire 3D Bloch sphere. This is an entirely different setup from what we propose in this work for TF-QKD, which always prepares states in a single mode (using phase-encoding) and performs slicing and matching for the different relative phases between pulses.

{\color{black}
The analysis and security implications are also vastly different for the two schemes: (1) For the decoy-state analysis, fully passive BB84 suffers from correlations between polarization and intensity distributions, which requires specific post-selection techniques to re-shape the two-dimensional polarization-intensity distribution before decoy-state analysis can be performed, while for fully passive TF-QKD there is only a single mode (polarization is fixed), and the decoy-state analysis is very straightforward for a signal following a one-dimensional intensity distribution, which is no different from that of the original passive decoy \cite{passivedecoy1} method. (2) For the security analysis, for fully passive BB84, the uncertainty in polarization in the key-generating Z basis cannot be considered as the result of a valid physical quantum channel, so we showed it can be considered as additional classical noise in post-processing (which does not increase privacy amplification needed). For fully passive TF-QKD, on the other hand, the uncertainty in encoded phase and intensity in the key-generating X basis always corresponds to a quantum channel, so we treat the system as having virtual ``local channels" which the signals from a perfect source must first pass through before entering the external real physical channel (and we compensate Z basis yields accordingly, to maintain basis independence of the local channels).

Lastly, compared with fully passive BB84, a big advantage fully passive TF-QKD offers is that it can eliminate side-channels from not only source modulators, but also detectors, enabling higher implementation security than either fully passive BB84 or active QKD.

Note that, alternatively, it is possible to implement a straightforward approach that offers similar protection against detector side-channels by directly duplicating the two-mode source in Ref. \cite{FullyPassiveThis} for each of Alice and Bob to form a fully passive measurement-device-independent (MDI) QKD setup (which is something we successfully implemented recently in Ref. \cite{passiveMDI1}, as well as reported in a concurrent paper \cite{passiveMDI2}). However, fully-passive MDI-QKD suffers from double the amount of sifting from both Alice and Bob, which, combined with increased inherent QBER from imperfectly prepared H and V states, results in a significant drop in key rate (about 3-4 orders-of-magnitude lower than that of active MDI-QKD). On the other hand, a fully passive TF-QKD scheme is based on summing up the key rate from all phase slice combinations, and it offers much better key rate (only about 1.2 orders-of-magnitude lower than active TF-QKD, which in itself already has significantly higher key rate than MDI-QKD). Overall, this makes fully passive TF-QKD the only viable choice at longer distances or for higher target key rates, while fully passive MDI-QKD actually complements the proposal at medium-to-close distances, due to its simplicity in experimental setup (no requirement on frequency stabilization and tracking) and equally-high implementation security.\\

Something to note for all fully-passive schemes (BB84, TF-QKD, MDI-QKD) is that, while their main advantage is removing source-side modulators, they do not grant the protocols source-independence. Indeed, Alice and Bob still need to protect their in-house laser sources and local detectors. However, these are generally much easier to protect than source modulators, since no signal is being sent \textit{out} from these detectors, and users can e.g. implement isolators to avoid backflowing signals from Eve. Moreover, local detectors operate at strong light levels, so even if any of Eve's signals leak through an isolator to the local detectors, they would be negligible compared to the signals.

Another assumption the fully-passive schemes build on is that the phase distributions coming from gain switched lasers are uniform and random, which is generally true for moderate repetition rates, as verified in e.g. \cite{PassiveExperiment01,mdiqkd_highspeed} in experiments. For very high repetition rates, though, the phase correlation could become more apparent between pulses, such as shown in Refs. \cite{phasecorrelation01,phasecorrelation02} (at 5GHz and 10GHz, respectively). This will result in not only imperfect phase randomization but also potentially decoy intensity and encoding correlations for fully passive schemes. In order to implement high-speed fully passive QKD schemes in the future, one might need to characterize the first-order and higher-order correlations \cite{phasecorrelation03} and potentially refine the security proof to incorporate the correlations, such as modifying upon the theories for intensity correlations \cite{phasecorrelation04} or imperfect phase randomization \cite{phasecorrelation05} (e.g. in the proof in Ref. \cite{phasecorrelation05}, the \textit{correlation} itself is considered as the effect of a local channel and attributed to Eve). Nonetheless, such correlations at high speed are not uniquely a challenge for passive TF-QKD, but a challenge for all fully passive and even active QKD schemes (e.g. Ref. \cite{phasecorrelation02} reported not just phase correlations but also polarization and intensity correlations for a high-speed active QKD system running at 5GHz). A security proof incorporating phase correlations will be a subject of future studies but is outside the scope of this work.

}

\section{The CAL Twin-Field QKD Protocol}

Here we present a brief recapitulation of the CAL TF-QKD protocol \cite{simpleTFQKD}, which the passive protocol in this work is based on. 

In the encoding basis X, Alice and Bob encode information by sending the states (for simplicity here we only show Alice's system, and Bob's system has the identical form but with suffixes $B$ and $b$):

\begin{equation}
	\ket{+}_{A}\ket{\alpha}_{a}+\ket{-}_{A}\ket{-\alpha}_{a}
\end{equation}

\noindent in the X basis.

In a virtual protocol, Alice sends cat states in the Z basis:

\begin{equation}
	\begin{aligned}
		&\ket{0}_{A}(\ket{\alpha}_{a}+\ket{-\alpha}_{a})/\sqrt{2}+\ket{1}_{A}(\ket{\alpha}_{a}-\ket{-\alpha}_{a})/\sqrt{2}\\
		=&\ket{0}_{A}\ket{C_0}_{a}+\ket{1}_{A}\ket{C_1}_{a},
	\end{aligned}
\end{equation}

\noindent where $\ket{C_0}_{a}$ and $\ket{C_1}_{a}$ are (unnormalized) cat states.

Alice and Bob measure their local qubits in X or Z bases. The phase error rate for the signal states is simply the bit error rate for the cat states. Moreover, using Cauchy-Schwarz inequality, any statistics for the cat states (which are superpositions of photon number states) can be upper-bounded by statistics of a mixture of photon number states. The upper bound for the phase error can be written as \cite{simpleTFQKD}:

\begin{equation}
	\begin{aligned}
		&Q_{k_c,k_d}^X e_{k_c,k_d}^Z \\
		&\leq \sum_{j=0,1}\left[\sum_{n_A,n_B=0}^{\infty} c_{m_A}^{A,(j)}c_{m_B}^{B,(j)} \sqrt{Y^Z_{{k_c},{k_d},m_A m_B}}\right]^2,\\
	\end{aligned}
\end{equation}

\noindent where $Y^Z_{{k_c},{k_d},m_A m_B}$ is the yield for the $n_A,n_B$ photon number state (here instead of using the notation $Y_{m_Am_B}$, in the following Appendices we use the more detailed notation from Ref. \cite{simpleTFQKD} and include the basis $Z$ and detection event $k_c,k_d$ information), and the coefficients for cat states are:

\begin{equation}
	\begin{aligned}
		c_{n_A}^{A,(j)} &= [(e^{-|\alpha_A|^2/2}{{\alpha_A^{n_A}}\over{\sqrt{n_A!}}})\\
		&+ (-1)^j \times (e^{-|\alpha_A|^2/2}{{(-\alpha_A)^{n_A}}\over{\sqrt{n_A!}}})]/2,\\
		c_{n_B}^{B,(j)} &= [(e^{-|\alpha_B|^2/2}{{\alpha_B^{n_B}}\over{\sqrt{n_B!}}})\\
		&+ (-1)^j \times (e^{-|\alpha_B|^2/2}{{(-\alpha_B)^{n_B}}\over{\sqrt{n_B!}}})]/2.\\
	\end{aligned}
\end{equation}

In practice, instead of using real cat states, in the Z testing basis, Alice and Bob prepare phase-randomized WCP states in various intensity settings, which can be used to construct a linear program to solve for $n_A,n_B$-photon yields $Y^Z_{{k_c},{k_d},n_A n_B}$ that are used above to calculate the phase error rate. 

The secure key rate is

\begin{equation}
	\begin{aligned}
		R_{k_c,k_d} &= Q^X_{k_c,k_d} [1-h_2(e^Z_{k_c,k_d})-f_e h_2 (E^X_{k_c,k_d})],\\
	\end{aligned}
\end{equation}

\noindent where $f_e$ is the error-correction efficiency and $E^X_{k_c,k_d}$ the observed quantum bit error rate in signal basis. The final key rate sums up the two detection patterns

\begin{equation}
	\begin{aligned}
		R &= R_{0,1} + R_{1,0}.
	\end{aligned}
\end{equation}

\section{Security Analysis for Passive TF-QKD}

As described in the main text, the finite size of the two local phase slices at each of Alice and Bob results in fluctuations in the intensity and phase of the signal states, which can potentially lead to a security loophole. 

To address this, we conceptually construct local quantum channels \cite{channel1,channel2,channel3} $\mathcal{E}_A$ and $\mathcal{E}_B$ inside Alice's and Bob's labs, as shown in main text Fig. 3, and attribute these channels to Eve (i.e. assuming they are part of the overall environmental noise/loss). For the signal basis, the only implication of this is higher QBER and consequently higher error-correction cost, which is already accounted for in the physical observables $E^X_{k_c,k_d}$.

The trickier part lies in the Z basis. We use the virtual protocol for CAL TF-QKD and assume Alice and Bob send perfect cat states in the Z basis. The cat states will now first pass through $\mathcal{E}_A$ and $\mathcal{E}_B$ before passing through the physical channel, which we can denote as $\mathcal{E}_E$. 

We can first apply Cauchy-Schwarz inequality at the perfect source, such that we only need to consider the statistics for Alice and Bob sending mixtures of $n_A,n_B$ photon number states first through $\mathcal{E}_A \otimes \mathcal{E}_B$ (whose effects are simply equivalent to photon losses, as shown in main text Fig. 4) and then through the external channel (whose effect is described by the yield $Y^Z_{{k_c},{k_d},m_A m_B}$, which can be obtained from decoy-state analysis).

The effect of $\mathcal{E}_A$, as shown in main text Fig. 4, can be described by a set of conditional probabilities $P_{\phi_{A1},\phi_{A2}}(m_A|n_A)$ of obtaining $m_A$ photons from Alice sending $n_A$ photons into the local channel (a similar process holds true for Bob):

\begin{equation}
	\begin{aligned}
		&P_{\phi_{A1},\phi_{A2}}(m_A|n_A) \\
		=& \left|{1\over{2!}} \sqrt{{{n_A!}\over{m_A!(n_A-m_A)!}}} \times \right. \\
		&\left[1+e^{i\pi/2 + (\phi_{A2}-\phi_{A1})}\right]^{m_A} \times \\
		&\left.\left[1+e^{i\pi/2 + (\phi_{A2}-\phi_{A1})}\right]^{(n_A-m_A)}\right|^2,
	\end{aligned}
\end{equation}

\noindent which depends on the phase difference between the upper and lower paths, $\phi_{A2}-\phi_{A1}$. The total yield can be obtained from Eq. 2 in the main text, which sums over all combinations of $m_A,m_B$ and integrates over the possible phase differences between $\phi_{A2}$ and $\phi_{A1}$.

{\color{black}
Note that, in Eq. 2 in the main text we have considered the slices to be in basis $l_{A1}=l_{A2}=l_{B1}=l_{B2}=0$ for simplicity, but in practice the slices can have any index combination, as shown in Fig. \ref{fig:localchannel1}. This simply results in a different integral region for the slices (and higher local photon losses):

\begin{equation}\label{eq:corrected_yield}
	\begin{aligned}
		&Y_{k_c,k_d,n_An_B}^{'Z} = \\
		&\int_{2l_{A1}\Delta_\phi-\Delta_\phi}^{2l_{A1}\Delta_\phi+\Delta_\phi} \int_{2l_{A2}\Delta_\phi-\Delta_\phi}^{2l_{A2}\Delta_\phi+\Delta_\phi} \int_{2l_{B1}\Delta_\phi-\Delta_\phi}^{2l_{B1}\Delta_\phi+\Delta_\phi} \int_{2l_{B2}\Delta_\phi-\Delta_\phi}^{2l_{B2}\Delta_\phi+\Delta_\phi} \\
		& \sum_{m_A=0}^{n_A}\sum_{m_B=0}^{n_B}P_{\phi_{A1},\phi_{A2}}(m_A|n_A)P_{\phi_{B1},\phi_{B2}}(m_B|n_B) \\
		&\times Y_{k_c,k_d,m_Am_B}^Z d\phi_{A1}d\phi_{A2}d\phi_{B1}d\phi_{B2},
	\end{aligned}
\end{equation}

\noindent where we have included the constant phase shifts due to the phase slice pair positions (bases) $\{l_{A1},l_{A2},l_{B1},l_{B2}\}$.\\

\begin{figure}[t]
	\includegraphics[scale=0.285]{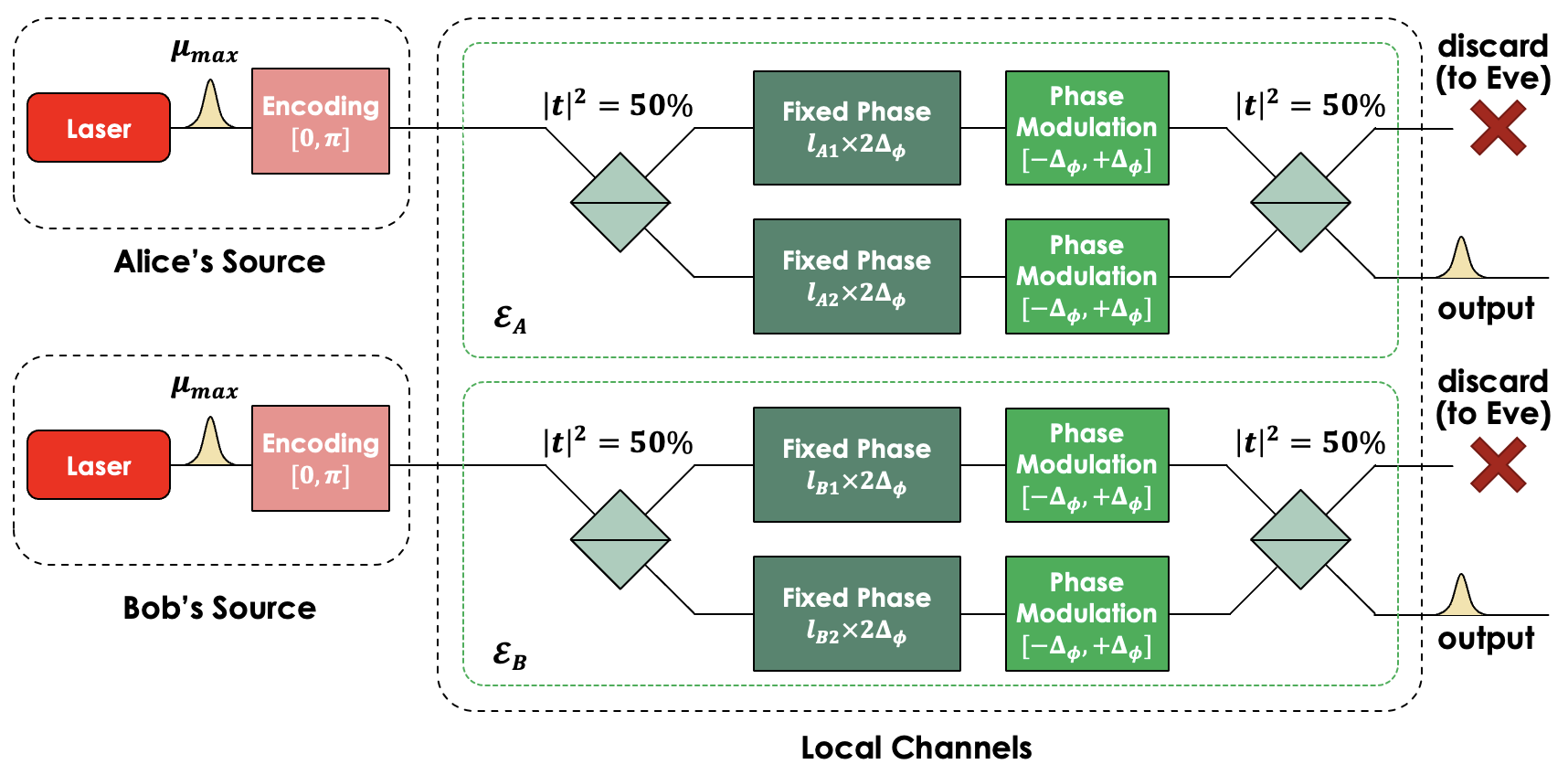}
	\caption{A more general version of Fig. 3 in the main text, where Alice's and Bob's slices are not in the default $\{0,0,0,0\}$ positions, but rather in bases $\{l_{A1},l_{A2},l_{B1},l_{B2}\}$, each of which adds a fixed phase shift of $l_i \times 2\Delta_\phi$. Compared to the default zero positions, these shifted slice positions will result in global phase shifts, increased average local loss, misalignments, and mismatched local losses between Alice and Bob, {\color{black}all of which are treated as part of the local channels $\mathcal{E}_A,\mathcal{E}_B$. Importantly, after correcting the local yields $Y_{k_c,k_d,n_An_B}^{'Z}$ to include effects of local losses, we now assume Alice and Bob send perfectly encoded signals from their sources, and $\mathcal{E}_A,\mathcal{E}_B$ are applied to their signals in both X and Z bases, i.e. the channels are basis-independent, which is why they can be attributed to Eve and considered part of the environment.}}
	\label{fig:localchannel1}
\end{figure}

	Here we also make a small clarification about the phrasing of ``attributing a channel to Eve" in the main text. We mean that (1) the settings of the local phase modulation and local loss are all public information known to Eve (which does not affect security, since the key is encoded in the $\ket{\alpha},\ket{-\alpha}$ states from the perfect laser source, not the phase fluctuations), and (2) the signal from the discarded output port from the beamsplitter is given to Eve, just like in an external lossy channel. This effectively allows us to consider the local loss/modulations as part of the environment in the security analysis.
	
	However, the assumptions do not imply that Eve can modify the settings inside Alice/Bob's lab. Here Alice and Bob still trust the values of their locally measured phase and losses, and they use them in Eq. \ref{eq:corrected_yield} to combine the physically observed yields with the local losses and obtain the corrected yield values (which contribute to the phase error rate estimation). This is natural, since from the beginning these settings are inside Alice's and Bob's local trusted labs, and pessimistically assuming that Eve can observe the settings or collect the output signal does not mean that Eve can change any of the settings.
	
	This is a common technique used in several prior art works reported in the literature, such as Refs. \cite{channel1,channel2,channel3} (particularly \cite{channel3}, which is explicitly in the QKD context and considers a mixing channel) where the quantum source sends perfectly encoded signals that go through a \textit{known} local channel/operation, which is assumed to be part of the external environment. More broadly speaking, earlier works, such as \cite{mdiqkd_experiment} employed a similar idea, where additional internal loss was added in the lab of the user (e.g. Bob, who was geographically closer to Charlie) in order to ensure overall system symmetry and good interference visibility. Physically, Alice and Bob were sending signals with asymmetric intensities out of their labs due to the additional local loss at Bob. The original protocol definition of MDI-QKD \cite{mdiqkd_practical} required symmetric source intensities, so Bob’s local loss was attributed to the environment (i.e. ``attributed to Eve”), such that the security analysis can still assume ``ideal" symmetric source intensities from Alice and Bob.
}

\section{Phase Slice Matching Strategy in the X Basis}

Here we describe in more detail the phase slice matching strategy that we propose in order to minimize the amount of discarded signals.

\subsection{Mismatches and Their Effect on Security}

\begin{figure*}[t]
	\includegraphics[scale=0.45]{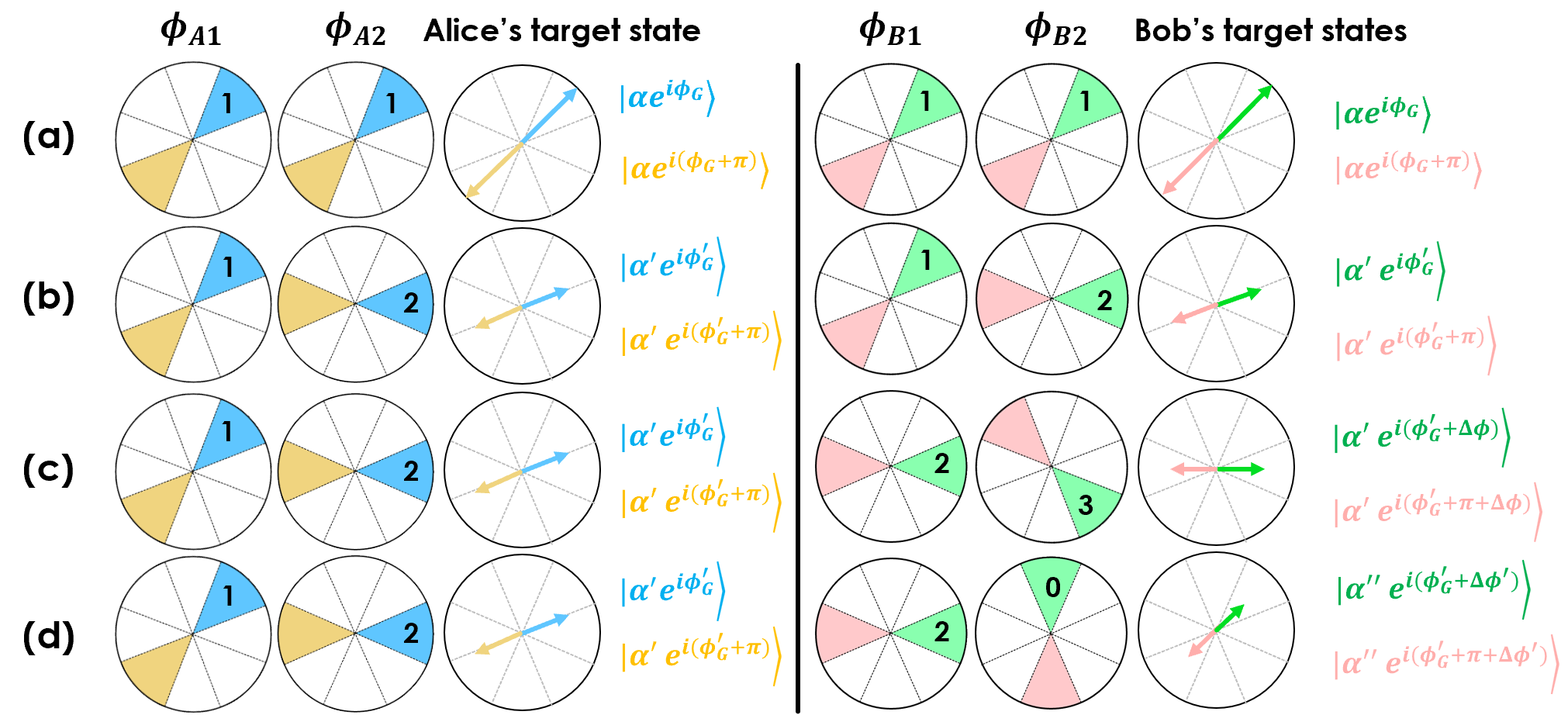}
	\caption{Types of mismatch between phase slices for sources A1, A2, B1, and B2. We also show the target states (the asymptotic limit of the state if the slice size is infinitely small) to be prepared. All types of mismatch only affect the key rate and will not require additionally modified security analysis. (a) Alice and Bob's phase reference points are not $l_i=0$ (i.e. not $\phi=0$); (b) Alice and Bob's local slices are not the same; (c) Alice and Bob's phase reference points are different; (d) The amount of mismatch between Alice and Bob's local slices are different. The above types of mismatch give us four degrees-of-freedom and can be used to describe any combination of $\{l_{A1},l_{A2},l_{B1},l_{B2}\}$.}
	\label{fig:a1}
\end{figure*} 

As mentioned in the main text, in the key-generating X basis, Alice and Bob measure the phases of their signals from sources A1, A2, B1, and B2 and divide the signals into phase slices indexed by $k_{A1},k_{A2},k_{B1}$ and $k_{B2}$. {\color{black}They then further pair up the opposite slices to represent the $0,1$ bits (i.e. corresponding to the $0,\pi$ phase encoding in active TF-QKD) for Alice and Bob. We can denote the key-generating ``basis" as $l_{A1},l_{A2},l_{B1}$ and $l_{B2}$, which are defined as

\begin{equation}
	\begin{aligned}
		l_{i} &= k_{i} \bmod (N/2) \;\;\;\; &(i\in\{A1,B1\}) \\
		l_{i} &= k_{i}  \;\;\;\; &(i\in\{A2,B2\})
	\end{aligned}	
\end{equation}

\noindent such that each set of $\{l_{A1},l_{A2},l_{B1},l_{B2}\}$ represents Alice choosing randomly between the pair $\{k_{A1}=l_{A1},k_{A2}=l_{A2}\}$ and the opposite pair $\{k_{A1}=l_{A1}+N/2,k_{A2}=(l_{A2}+N/2) \bmod N\}$, where both local slices are shifted by phase $\pi$. Bob similarly chooses his $\{k_{B1},k_{B2}\}$ among two opposite pairs differing by a $\pi$ phase shift. Each set of ``basis" $\{l_{A1},l_{A2},l_{B1},l_{B2}\}$ therefore represents four possible patterns of $\{k_{A1},k_{A2},k_{B1},k_{B2}\}$. There are $N^4/4$ sets of basis in total.

When performing key generation, only the phase slice pair (basis) information $l_i$ but not the actual phase slice $k_i$ is publicly announced by Alice and Bob. Each combination of $\{l_{A1},l_{A2},l_{B1},l_{B2}\}$ is separately used to perform privacy amplification, error-correction and generate key.}

Compared to the default case as shown in main text Fig. 2 (a), where Alice and Bob pick $\{l_{A1}=0,l_{A2}=0,l_{B1}=0,l_{B2}=0\}$, i.e. the same local slice and both use the same phase reference, in the general case there are four possible types of mismatches, which we show in Fig. \ref{fig:a1}. We will proceed to explain that all of these mismatches are already included in the existing TF-QKD protocol framework and do not require additional revision to the security analysis:

\begin{enumerate}
	\item As shown in Fig. \ref{fig:a1} (a), the global phase reference for Alice (and also Bob) is not zero. If Eve does not interfere, this will not have any effect on either bit error rate (since it generates identical statistics at Charlie, who only compares phase difference between Alice and Bob) or the phase error rate (since phase shifts do not affect the Fock state yields), leading to the same secure key rate as slices with phase reference point at zero. If Eve adds some global-phase-dependent operations, they will only potentially decrease the key rate but not affect the security (as they will always be considered external channel noise). We will discuss a bit further on how to make use of the rotational symmetry of the global phase to combine some signals in the following subsection D. 2.
	
	\item As shown in Fig. \ref{fig:a1} (b), the local slices are mismatched at Alice (and also Bob). This again results in a shifted phase reference (which has no effect on the key rate so long as the amount of shifting is the same for Alice and Bob) as well as a smaller average output signal intensity. {\color{black}As shown in Fig. \ref{fig:localchannel1} in Appendix C, a fixed phase shift, e.g. $l_{A2}-l_{A1}$, simply changes the transmittance of the local channel $\mathcal{E}_A$. In our source model, we can always assume the virtual sources to be perfectly preparing signals $\ket{\pm \sqrt{\mu_{max}}}$, while all attenuations to the signal intensity are attributed to $\mathcal{E}_A$ and considered as part of channel loss. One effect is that the final output intensity in the X basis becomes smaller, and as a result the X basis gain also decreases. The other effect is that the corrected yields will also be adjusted, due to the increased local loss.}
	
	\item As shown in Fig. \ref{fig:a1} (c), Alice and Bob's phase reference points can be different. This is equivalent to a channel misalignment and will result in higher QBER in the X basis and thus lower key rate, but will not affect the security.
	
	\item As shown in Fig. \ref{fig:a1} (d), the amount of mismatch between Alice's and Bob's local slices can be different. This on the one hand results in different phase reference points (same as mentioned above in point 3) and on the other hand {\color{black}results in different levels of local channel losses for Alice and Bob. Here, again, in our model we assume the virtual sources to be still sending $\ket{\pm \sqrt{\mu_{max}}}$, but as discussed in Ref. \cite{asymTFQKD}, asymmetric channel transmittances simply result in worse visibility (hence higher QBER). The security, again, will not be affected.}
\end{enumerate}

{\color{black}

These four degrees-of-freedom (representing four types of mismatches) $x_1=(l_{A1}+l_{A2})/2$, $x_2=l_{A2}-l_{A1}$, $x_3=(l_{A1}+l_{A2})/2 - (l_{B1}+l_{B2})/2$, and $x_4=(l_{B2}-l_{B1})-(l_{A2}-l_{A1})$, are sufficient to represent all combinations of $\{l_{A1},l_{A2},l_{B1},l_{B2}\}$, meaning that the use of any such combination will be a legitimate setup for CAL TF-QKD (and the mismatches will be considered as a global phase shift, a local intensity modulation, a remote phase misalignment, and a remote intensity mismatch). Therefore, we can simply add up the non-zero key rates, in the form of Eq. 1 in the main text:

\begin{equation}\label{eq:rate}
	\begin{aligned}
		R =& {4 \over {N^4}} \sum_{l_{A1}=0}^{(N/2)-1} \sum_{l_{A2}=0}^{N-1} \sum_{l_{B1}=0}^{(N/2)-1} \sum_{l_{B2}=0}^{N-1}\\
		&[\max(0,R_{0,1} (l_{A1},l_{A2},l_{B1},l_{B2})) \\
		&+ \max(0,R_{1,0} (l_{A1},l_{A2},l_{B1},l_{B2}))],
	\end{aligned}
\end{equation}

Note that here there are $N^4/4$ unique sets of $\{l_{A1},l_{A2},l_{B1},l_{B2}\}$, and each of the phases $\phi_{A1},\phi_{A2},\phi_{B1},\phi_{B2}$ (which are assumed to follow uniform distributions between $[0,2\pi)$) have ${4 /{N^4}}$ probability to fall within a particular set.

\subsection{Using Additional Symmetry}

\begin{figure}[t]
	\includegraphics[scale=0.21]{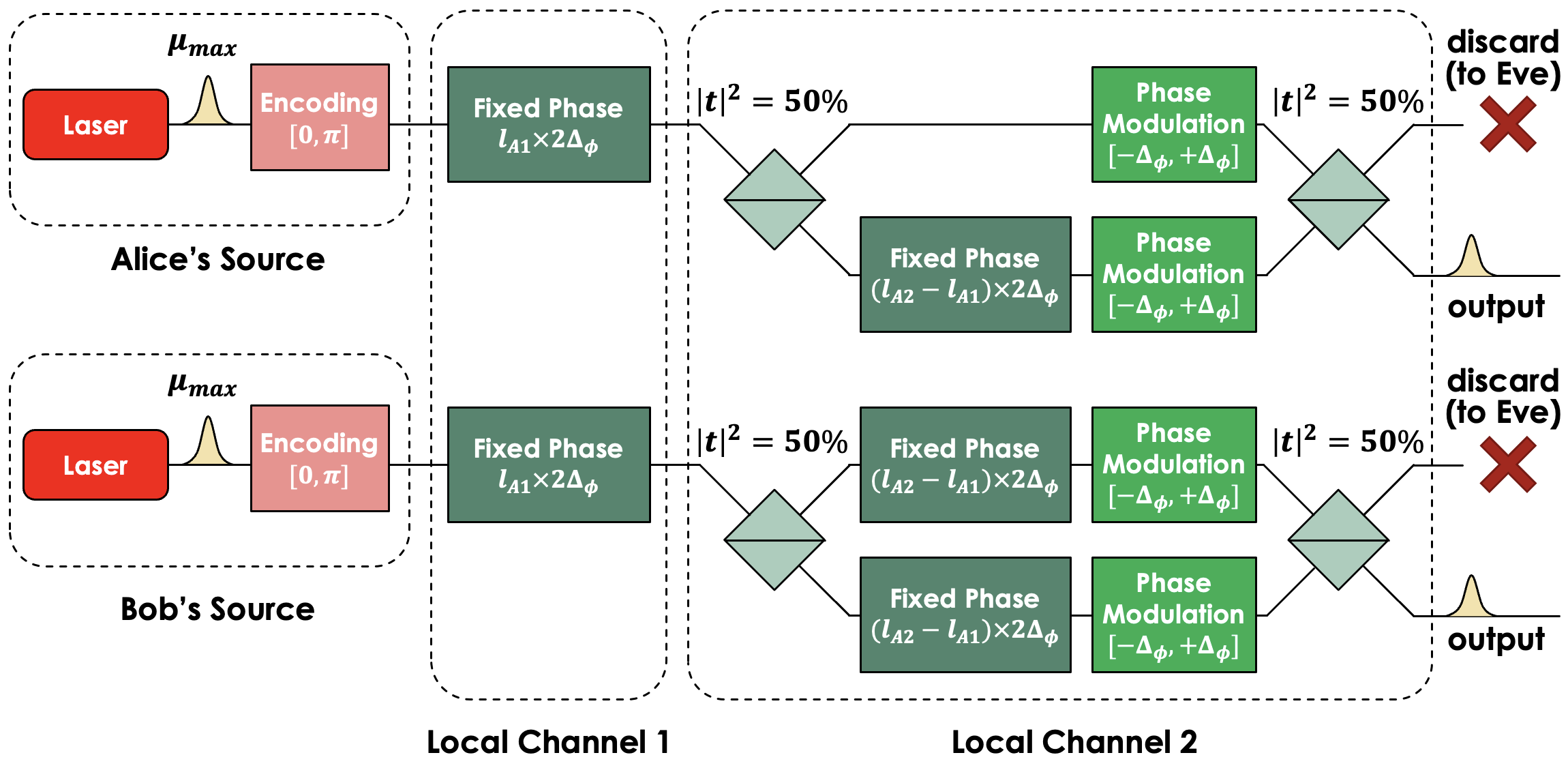}
	\caption{Using rotational symmetry of phase slice positions. Here we can extract a common $l_{A1} \times 2\Delta_{\phi}$ out of the fixed phases, such that it can be considered as another separate local channel that rotates Alice's and Bob's signals by a common global phase of $2l_{A1} \Delta_{\phi}$. Alice and Bob divide signals based on $\{l_{A2}-l_{A1},l_{B1}-l_{A1},l_{B2}-l_{A1}\}$ into separate rounds of privacy amplification, but signals with different global phase $2l_{A1}\Delta_{\phi}$ are grouped into the same round. Importantly, both local channels 1 and 2 in the figure are considered as being attributed to Eve. If Eve does not interfere, in a normal experimental setting, the varying global phase in channel 1 has no effect on observed statistics, while if Eve does add global-phase-dependent noise, it will only increase QBER but will not affect any security assumptions, since the phase rotations and any phase-dependent attacks are all considered as external channel noise.}
	\label{fig:symmetry}
\end{figure} 

Additionally, to simplify the calculations, we can utilize the rotational symmetry and revise the key rate formula as:

\begin{equation}\label{eq:rate_symmetry}
	\begin{aligned}
		R =& {2 \over {N^3}} \sum_{l_{A2}'=0}^{N-1} \sum_{l_{B1}'=0}^{(N/2)-1} \sum_{l_{B2}'=0}^{N-1}\\
		&[\max(0,R_{0,1} (0,l_{A2}',l_{B1}',l_{B2}')) \\
		&+ \max(0,R_{1,0} (0,l_{A2}',l_{B1}',l_{B2}'))],
	\end{aligned}
\end{equation}

\noindent where $l_{A2}'=l_{A2}-l_{A1},l_{B1}'=l_{B1}-l_{A1},l_{B2}'=l_{B2}-l_{A1}$\footnote{Note that in fact $l'_{B1}$ ranges from $-l_{A1}$ to $N/2-1-l_{A1}$, but since a slice pair with position $-l_i$ simply means $N/2-l_i$, the range is still the same $0$ to $N/2-1$.}. In other words, we combine the key rate for signals that have the same $l_{A2}-l_{A1},l_{B1}-l_{A1},l_{B2}-l_{A1}$, regardless of whether they might have different global phases (determined by $l_{A1}$). Geometrically, this is saying that we can combine the signals from all phase slice patterns $\{l_{A1},l_{A2},l_{B1},l_{B2}\}$ that differ only by a rotation for one single round of privacy amplification, error correction, and key generation.

Such combination of signals does not affect security. The reason is that, as shown in Fig. \ref{fig:symmetry}, the global phase rotation $l_{A1} \times 2\Delta_{\phi}$ can be considered as a local channel, too, while we can still assume Alice and Bob to send the perfect signals $\ket{\pm \sqrt{\mu_{max}}}$ with no phase rotation nor intensity modulation. If Eve is not making an attack, such phase rotations will not change either the observables (gain and QBER) in the X basis or the Fock state yields in the Z basis, so it is natural to combine the signals for key generation. If Eve does make a global-phase-dependent attack, both the phase rotation and the attack will be considered as a part of the external channel, so the most Eve can do is just to increase the QBER and phase error rate, but not to affect the security of the protocol. \footnote{Of course, if, say, the QBER does vary depending on the global phase, due to some special channel properties or Eve's attack, combining signals with various global phases is indeed less optimal than calculating the key rate for each case separately, at least in the asymptotic case. This is because of the concavity of the binary entropy function $h_2$, such that $\langle h_2(QBER) \rangle \leq h_2(\langle QBER \rangle)$ based on Jensen's inequality. However, this is only about how to obtain the optimal key rate, but, importantly, the lower bounds on the key rates using either combined or separated signals are both valid and secure.}

Conceptually, this is no different from collecting data from a long day of QKD experiment and using all data for a single round of post-processing and key generation, while Eve does some time-dependent attack that introduces different levels of noise throughout the day (or maybe the inherent channel noise/loss changes throughout the experiment, such as in satellite QKD). Again, such a time-dependent attack/noise in the experimental data might increase the overall average QBER or phase error rate (thus potentially decreasing key rate or even causing denial-of-service), but at the end of the day, it will not affect the security of the final estimated lower bound on the key rate. 

Note that, in the asymptotic case with infinite data, utilizing rotational symmetry and combining the above-said patterns into the same round of privacy amplification makes no difference to the asymptotic key rate, and it is mainly done to help simplify post-processing and reduce computation time (both in the simulation and in the analysis of experimental data). However, as we will discuss again in Appendix I, such combination of data using symmetry will provide an additional benefit of helping reduce statistical fluctuation in the finite-size scenario.
}

\section{Decoy Analysis in the Z Basis}

In this section we describe the models of decoy-state analysis we used in the Z basis when performing the simulations shown in main text Fig. 5, including that of active TF-QKD, passive TF-QKD with infinite decoys, as well as passive TF-QKD with a finite number of decoys. Note that, so far we only focus on the scenario of infinite signals (some additional discussions on the scenario with finite data size are included later in Appendix I). Here in this section, when using terms including ``asymptotic" and ``finite", we are referring to the number of decoy-state settings. 

\textit{1. Active TF-QKD}: we simply follow the practical protocol 3 from \cite{simpleTFQKD}, i.e. CAL TF-QKD with signal states of phases $\{0,\pi\}$ in the X basis, and phase-randomized decoy states with various intensity settings in the Z basis. In our simulation, we consider the asymptotic case with infinite decoy settings, i.e. Alice and Bob have perfect knowledge of all the photon-number yields $Y_{m_Am_B}$. We calculate the phase error rate simply using the theoretical values for the yields. The full form of yields in the presence of channel polarization misalignment is shown in \cite{simpleTFQKD}. Here for simplicity we consider no polarization misalignment, in which case the yields can be simplified into:

\begin{equation}\label{eq:theory_yield}
	\begin{aligned}
		Y_{k_c,k_d,m_Am_B}^{Z,p_d=0}=&\sum_{k=0}^{m_A}{{m_A}\choose{k}}\sum_{l=0}^{m_B}{{m_B}\choose{l}}\\
		&\times {{\eta^{k+l}(1-\eta)^{m_A+m_B-k-l}(k+l)!}\over{2^{k+l}k!l!}}\\
		&- (1-\eta)^{m_A+m_B},
	\end{aligned}
\end{equation}

\noindent where $\eta$ is the channel loss between Alice (Bob) and Charlie. In the presence of dark count probability $p_d$, the corrected yields are

\begin{equation}\label{eq:theory_yield2}
	\begin{aligned}
		Y_{k_c,k_d,m_Am_B}^Z=&(1-p_d)\left[p_d(1-\eta)^{m_A+m_B}+Y_{k_c,k_d,m_Am_B}^{Z,p_d=0}\right].
	\end{aligned}
\end{equation}\\

\textit{2. Passive TF-QKD with infinite decoys}: For passive TF-QKD, we first consider the case of infinite decoy settings, where we again assume that we have perfect knowledge of all photon-number yields $Y_{m_Am_B}$ and use Eqs. \ref{eq:theory_yield} and \ref{eq:theory_yield2}. The external channel photon-number yields $Y_{m_Am_B}$ are further used in calculating the corrected yields $Y_{n_An_B}$ in Eq. 2 in the main text when including the effect of local channels in Alice's and Bob's labs in the passive TF-QKD security analysis. The corrected yields are then combined with passive signal state statistics to calculate the key rate.

There is actually a slight caveat here: for passive TF-QKD, since there is no active basis-switching and the same signal is used for X and Z bases (just with different post-selection/post-processing strategies), the decoy intensities are limited to the range of $\mu_A\in [0,\mu_{max}]$ and $\mu_B\in[0,\mu_{max}]$ and are capped by $\mu_{max}$, which needs to be small because the phase error rate estimation favors smaller signal intensities (which affect the cat state coefficients). This is different from the ideal infinite-decoy case for active TF-QKD, where the decoy intensities can in principle go up to infinity. 

However, we still have the freedom to choose as many divisions (each of which can also be infinitesimally small, such that it is equivalent to using a fixed intensity value, instead of a range of intensities) within $[0,\mu_{max}]$ as we want. This means that, in theory, we can still get an infinite number of linearly independent equations, which are sufficient to solve any arbitrarily large numbers of variables $Y_{m_Am_B}$ accurately, which justifies our usage of theoretical values of $Y_{m_Am_B}$ when calculating the key rate for infinite-decoy passive TF-QKD.

In the simulation in main text Fig. 5, we optimized $\mu_{max}$, with optimal values falling between $0.016-0.03$.\\

\textit{3. Passive TF-QKD with a finite number of decoy settings}: Lastly, for practical passive TF-QKD using a finite number of decoy settings, we need to divide up the map of $\mu_A,\mu_B$ into decoy regions (such as the ones shown in main text Fig. 2 (b)) and perform passive decoy-state analysis. Each region functions as one decoy setting, where we calculate the gain and the photon number distribution by integrating over the given region, similar to what is performed in passive decoy-state BB84 protocol \cite{passivedecoy1,passivedecoy2} \footnote{Note that this is different from the decoy-state analysis in fully passive BB84 \cite{FullyPassiveThis,FullyPassiveAlternative}, where the analysis is significantly more complicated since polarization (and by extension the yield and QBER) also depends on $S_{ij}$. Here for passive TF-QKD, the decoy-state analysis is very straightforward: the only difference from the active case is that the photon number distribution is averaged over $S_{ij}$, but, importantly, the yields $Y_{k_c,k_d,m_Am_B}^Z$ do not depend on $\mu_A$ and $\mu_B$ and are independent of the integral.}. The average photon number distribution is:

\begin{equation}
	\begin{aligned}
		\langle P^Z(m_A,m_B)\rangle &= {1\over{P_{S_{ij}}}} \iint_{S_{ij}} \\
		&P_{Poisson}(\mu_A,m_A)P_{Poisson}(\mu_B,m_B)\\
		& P_{int}(\mu_A,\mu_B) d\mu_Ad\mu_B,
	\end{aligned}
\end{equation}

For instance for a grid of $3\times 3$ square regions, the distribution for each setting $(a,b)$ (where $a,b=1,2,3$ and square regions are used) is

\begin{equation}
	\begin{aligned}
		\langle P^Z(m_A,m_B) \rangle &= {1\over{P_{S_{ij}}}} \int_{(a-1)\mu_{max}/3}^{a\mu_{max}/3} \int_{(b-1)\mu_{max}/3}^{b\mu_{max}/3} \\
		&P_{Poisson}(\mu_A,m_A)P_{Poisson}(\mu_B,m_B)\\
		& P_{int}(\mu_A,\mu_B) d\mu_Ad\mu_B,
	\end{aligned}
\end{equation}

\noindent where the Poissonian distribution is

\begin{equation}
	P_{Poisson}(\mu,n) = e^{-\mu} {{\mu^n}\over {n!}}.
\end{equation}

Once we have the set of observables $\{Q_{k_c,k_d,i,j}^Z\}$, as well as the photon number distributions for each $(m_A,m_B)$, we can list out the linear equations

\begin{equation}
	Q_{k_c,k_d,i,j}^Z = \sum_{m_A,m_B} \langle P^Z(m_A,m_B) \rangle Y^Z_{{k_c},{k_d},m_A m_B},
\end{equation}

\noindent from which we can estimate the upper bounds of photon-number yields $Y^Z_{{k_c},{k_d},m_A m_B}$ for the physical external channels, which can be used to calculate the corrected yields in main text Eq. 2 (including the effects of the local virtual channels) and further bound the phase error rate $e_{k_c,k_d}^Z$.\\

\begin{figure}[t]
	\includegraphics[scale=0.23]{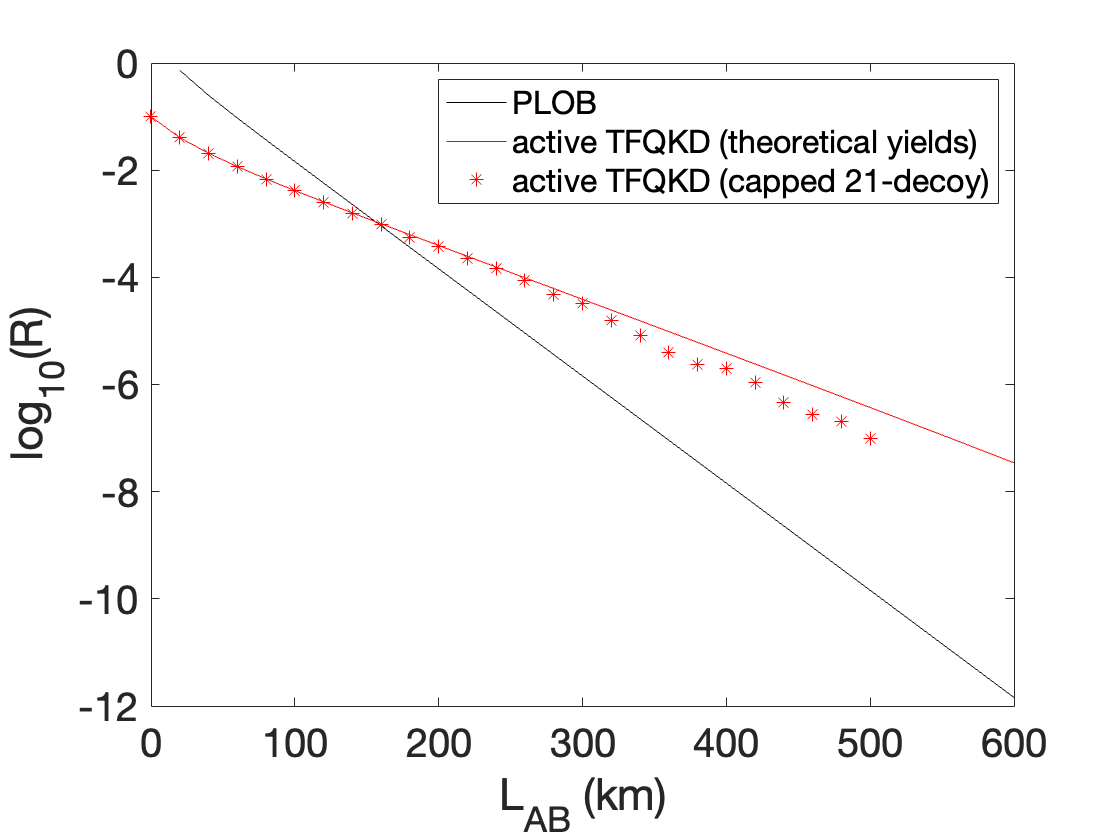}
	\caption{Comparison of key rates for active TF-QKD calculated with infinite number of decoys using theoretical values of yields and with decoy settings capped with a maximum intensity of $\mu_{max} \approx 0.1$ and solved with linear programming. We optimize the signal intensity for the case with theoretical yields and also use the value as the maximum intensity $\mu_{max}$ for the case with capped intensities. For the linear program, we choose as many number of decoys as numerically feasible in order to approach the infinite-decoy case (here we use $linspace(0,0.05,1)\times \mu_{max}$, where $linspace$ is an operator that returns an array of equidistant values between $0$ and $1$ with a step size of $0.05$, i.e. using $21\times 21$ decoy settings for Alice and Bob in total). No dark count or misalignment is introduced and the channels only contain loss. As can be seen, even with a large number of decoy settings and no noise, the key rate solved with linear programming is visibly lower at longer distances, illustrating the numerical instability caused by having small coefficients in the linear program, which is a side-effect of using small decoy intensities, despite that, theoretically, having capped intensities should not affect one's bound on the yields so long as one has an infinite number of decoy settings.}
	\label{fig:a2}
\end{figure} 
 
Note that, as we mentioned above, the decoy intensities being capped to a small value does not fundamentally limit the asymptotic upper bound for the key rate of passive TF-QKD, since we can in principle still choose an infinite number of infinitesimally small decoy regions, which form linearly independent constraints, and solve for the yields with arbitrarily high precision. 

However, in practice, the smaller decoy intensity values may still cause some numerical instabilities for the linear solvers, due to the coefficients being small and neighboring linear equations having coefficient values that are rather close to each other. Small numerical inaccuracies might even sometimes inadvertently cause some sets of linear constraints to be contradictory, making the linear program infeasible for some $Y_{m_Am_B}$ (in which case we have to upper-bound them to one, increasing the phase error rate and decreasing the key rate). We can illustrate this problem in Fig. \ref{fig:a2} where we compare an ideal case of active TF-QKD with infinite decoy settings and unlimited intensity range (i.e. yields simply approach theoretical values) versus the same active protocol with decoy intensities capped to a small value and solved with linear programming, where we give it 21 decoy intensities, which should theoretically be very close to the infinite-decoy scenario. However, we see a visible drop for the case with capped intensities, simply due to numerical inaccuracies.

In the simulation in main text Fig. 5, the signal intensity, as well as the values $t_{decoy1}$ and $t_{decoy2}$ (which we allow to be adjustable values, rather than $2/3$ and $1/3$) that define the grids for the decoy regions $[0,t_{decoy2}\mu_{max})$, $[t_{decoy2}\mu_{max},t_{decoy1}\mu_{max})$, $[t_{decoy1}\mu_{max},\mu_{max}]$ are all optimized. In the asymptotic scenario, decoy analysis always favors as small ``vacuum" and ``weak" decoy setting intensities as possible, so we just set $t_{decoy2}$ to a reasonable $0.1$ and optimize $\mu_{max}$ and $t_{decoy1}$. Optimal $\mu_{max}$ falls within $0.006-0.03$ and optimal $t_{decoy1}$ varies between $0.12-0.93$.

As discussed above, our passive TF-QKD scheme with three decoy settings also suffers from such numerical challenges, meaning that the numerically bounded key rate in main text Fig. 5 is slightly lower than what it could have theoretically been. Note that, however, the above problem is not a theoretical limitation but just a numerical one, and e.g. scaling methods for the linear program can alleviate the problem to an extent (similar to the numerical problem and the methods to alleviate it as mentioned in Ref. \cite{multiphoton_interference}). 

Also, experimentally, the smaller decoy intensities put a higher requirement on the accuracy of intensity measurement, but since the intensity measurement is performed on classical strong light, in principle we can have arbitrarily high accuracy for the measurements (albeit in reality this will be limited by the intensity of the laser diode, as well as the resolution and noise of the photodiodes used for detection). Again, this is not a theoretical limitation on our use of small decoy intensities, but rather an engineering one that can be lifted or at least alleviated with better equipment or experimental design.

{\color{black}
	Lastly, the smaller decoy intensities mean that the strongest decoy state is never stronger than the signal state in passive TF-QKD. This is not a problem asymptotically, but can be a problem for some finite-size analysis model, as we will discuss in more detail in Appendix I. However, as we will show in Appendix I, we can propose theoretical workarounds to reduce the signal intensities in the ``virtual source" before entering Alice's and Bob's local channels, such that signal intensities are still weaker than the strongest decoy intensity. 
}

\section{Source and Channel Model}

\subsection{Simulation of Observables}

Firstly, we consider the coding phase. Alice and Bob observe the phases of their local sources $\phi_{A1},\phi_{A2},\phi_{B1}$ and $\phi_{B2}$. The signals are binned into phase slices $k_{A1},k_{A2},k_{B1},k_{B2}$, which are further paired up to form sets of ``bases" indexed by $l_{A1},l_{A2},l_{B1}$ and $l_{B2}$. Within each slice, the phase is distributed within a range of $[-\Delta_\phi,\Delta_\phi]$ where $\Delta_\phi=\pi/N$. 

Here we first consider the statistics as a function of a given set of exact phases $\phi_{A1},\phi_{A2},\phi_{B1}$ and $\phi_{B2}$, which can later be integrated over domains that are dependent on the phase slice choices.

In main text Fig. 1 (a), we can denote Alice's (Bob's) input ports of the rightmost beam splitter as A1,A2 (B1,B2), and output ports as A3,A4 (B1,B4). The amplitudes of signals at A3 and B3 are:

\begin{equation}\label{eq:source_interference}
	\begin{aligned}
		\alpha_{A3} &= {\alpha_0 \over \sqrt{2}} e^{i\phi_{A1}} + i {\alpha_0 \over \sqrt{2}} e^{i(\phi_{A2}-\pi/2)},\\
		\alpha_{B3} &= {\alpha_0 \over \sqrt{2}} e^{i\phi_{B1}} + i {\alpha_0 \over \sqrt{2}} e^{i(\phi_{B2}-\pi/2)},
	\end{aligned}
\end{equation}

\noindent where $\alpha_0 = \sqrt{\mu_0} = \sqrt{u_{\max}/2}$ for each input signal, and the output amplitudes are complex numbers (i.e. vectors in complex space). We can convert $\alpha_{A3},\alpha_{B3}$ from Cartesian to polar coordinate in the complex space:

\begin{equation}
	\begin{aligned}
		[Re(\alpha_{A3}),Im(\alpha_{A3})] &\rightarrow [\alpha_A,\phi_A], \\
		[Re(\alpha_{B3}),Im(\alpha_{B3})] &\rightarrow [\alpha_B,\phi_B],
	\end{aligned}
\end{equation}

\noindent where we can now treat Alice's and Bob's output states as if they were two coherent light sources of amplitudes and phases $(\alpha_A,\phi_A),(\alpha_B,\phi_B)$. Let $\eta$ be the channel loss between Alice and Charlie (and Bob and Charlie, assuming symmetric channels). The output ports $C,D$ at Charlie have amplitudes:

\begin{equation}
	\begin{aligned}
		\alpha_{C} &= {\alpha_A \sqrt{\eta} \over \sqrt{2}} e^{i\phi_{A}} + i {\alpha_B \sqrt{\eta} \over \sqrt{2}} e^{i(\phi_{B}-\pi/2)},\\
		\alpha_{D} &= i {\alpha_A \sqrt{\eta} \over \sqrt{2}} e^{i\phi_{A}} + {\alpha_B \sqrt{\eta} \over \sqrt{2}} e^{i(\phi_{B}-\pi/2)}.
	\end{aligned}
\end{equation}

For a given detector with a coherent light of amplitude $\alpha$ arriving at it, we can calculate the click probability:

\begin{equation}
	\begin{aligned}
		P_{\text{click}} = 1 - e^{-|\alpha|^2} (1-p_d),
	\end{aligned}
\end{equation}

\noindent where $p_d$ is the dark count probability. Combining the data from detectors $C$ and $D$ gives us the singles probability 

\begin{equation}
	\begin{aligned}
		P_{k_c,k_d}^X(\phi_{A1},\phi_{A2},\phi_{B1},\phi_{B2})&=(P_{\text{click},C})^{k_c}(1-P_{\text{click},D})^{k_c}\\
		&\times (P_{\text{click},D})^{k_d}(1-P_{\text{click},C})^{k_d}
	\end{aligned}
\end{equation}

\noindent where $k_c,k_d$ can be bits $0,1$ or $1,0$. Note that this click probability is solely a function of the four local phases, $(\phi_{A1},\phi_{A2},\phi_{B1},\phi_{B2})$.

Next, considering the phase slices chosen, the actual observed data is a 4-dimensional integral of the click statistics over the slices.

\begin{equation*}
	\begin{aligned}
		Q_{k_c,k_d,+}^X &= {1\over{(2\Delta_\phi)^4}} \int_{2l_{A1}\Delta_\phi-\Delta_\phi}^{2l_{A1}\Delta_\phi+\Delta_\phi}\int_{2l_{A2}\Delta_\phi-\Delta_\phi}^{2l_{A2}\Delta_\phi+\Delta_\phi} \\
		&\int_{2l_{B1}\Delta_\phi+\phi_E-\Delta_\phi}^{2l_{B1}\Delta_\phi+\phi_E+\Delta_\phi}\int_{2l_{B2}\Delta_\phi+\phi_E-\Delta_\phi}^{2l_{B2}\Delta_\phi+\phi_E+\Delta_\phi}\\
		&P_{k_c,k_d}^X(\phi_{A1},\phi_{A2},\phi_{B1},\phi_{B2}) d\phi_{A1}\phi_{A2}\phi_{B1}\phi_{B2}, \\
		Q_{k_c,k_d,-}^X &= {1\over{(2\Delta_\phi)^4}} \int_{2l_{A1}\Delta_\phi-\Delta_\phi}^{2l_{A1}\Delta_\phi+\Delta_\phi}\int_{2l_{A2}\Delta_\phi-\Delta_\phi}^{2l_{A2}\Delta_\phi+\Delta_\phi} \\
		&\int_{2l_{B1}\Delta_\phi+\phi_E+\pi-\Delta_\phi}^{2l_{B1}\Delta_\phi+\phi_E+\pi+\Delta_\phi}\int_{2l_{B2}\Delta_\phi+\phi_E+\pi-\Delta_\phi}^{2l_{B2}\Delta_\phi+\phi_E+\pi+\Delta_\phi}\\
		&P_{k_c,k_d}^X(\phi_{A1},\phi_{A2},\phi_{B1},\phi_{B2}) d\phi_{A1}\phi_{A2}\phi_{B1}\phi_{B2},
	\end{aligned}
\end{equation*}

\noindent where the two parities $+,-$ depend on the encoding bits that are sent (and how Alice and Bob determine which pairs correspond to 0 or 1 bits). \footnote{For simplicity, here in the simulation model we assume a channel whose statistics do not depend on global phase, hence Alice and Bob sending 00 and 11, or alternatively 10 and 01, results in the same statistics, and each parity case only needs to be simulated once. If we want to express a global-phase-sensitive channel, we can simply simulate all four combinations of signals corresponding to the 00,01,10,11 bits, instead of the even/odd parities.}. Here we characterize the misalignment between Alice and Bob as $\phi_E$.

One thing worth noting here is that, for our passive TF-QKD protocol, Alice and Bob generate key from \textit{all} combinations of slices. Since the misalignment rotates all of Bob's slices, there has to be \textit{some} slice position on Bob's side that minimizes the misalignment, so Bob can simply rotate his slice indexing to start from there. Therefore, a simple phase misalignment $\phi_E$ would affect the overall key rate very little (only the portion of $|\phi_E|<\Delta_\phi$ will have any effect). This is an additional benefit of using the passive encoding scheme we propose.\footnote{By the way, as mentioned in the main text, such an effect of resistance against misalignment is similar to that observed in phase-matching (PM) QKD, which also uses phase slices on a circumference and can thus simply rotate the slice index to minimize the misalignment.}

Depending on how Alice and Bob choose the 0 vs 1 bits (ideally, the smaller one of $Q_{k_c,k_d,+}^X ,Q_{k_c,k_d,-}^X$ should be defined as the error), the QBER can be written as:

\begin{equation}
	\begin{aligned}
		Q_{k_c,k_d}^X &= (Q_{k_c,k_d,+}^X +Q_{k_c,k_d,-}^X ), \\
		E_{k_c,k_d}^X &= min(Q_{k_c,k_d,+}^X ,Q_{k_c,k_d,-}^X) / (Q_{k_c,k_d,+}^X +Q_{k_c,k_d,-}^X).
	\end{aligned}
\end{equation}\\

Secondly, we consider the testing phase (decoy states). Here, we consider Alice and Bob to have two phase-randomized sources with arbitrary intensities $(\mu_A,\mu_B)$, each of which takes a value between $[0,\mu_{max}]$. Fundamentally, the values are determined by $|\phi_{A1}-\phi_{A2}|$ and $|\phi_{B1}-\phi_{B2}|$, but for convenience we convert the degrees-of-freedom into $(\mu_A,\mu_B)$ directly. Note that, while the phases are uniformly distributed over $[0,2\pi)$, the corresponding intensities satisfy the probability distribution \cite{FullyPassiveThis} of

\begin{equation}\label{eq:pint}
	p_{int}^{2D}(\mu_A,\mu_B) = {1 \over {\pi^2 \sqrt{\mu_A(\mu_{max}-\mu_A)}}\sqrt{\mu_B(\mu_{max}-\mu_B)}}.
\end{equation}

This probability distribution integrates to 1 over the region $[0,\mu_{max}]\times [0,\mu_{max}]$. Note that the intensities $\mu_A,\mu_B$, the random relative phase $\phi_{AB}$ between Alice and Bob, together with a random global phase (which ensures the validity of the photon-number assumption, and cannot be used for post-selection), contain the same degrees-of-freedom as $(\phi_{A1},\phi_{A2},\phi_{B1},\phi_{B2})$.

For any given set of $(\mu_A,\mu_B,\phi_{AB})$, the amplitudes arriving at detectors $C,D$ are

\begin{equation}
	\begin{aligned}
		\alpha_{C} &= {\sqrt{\eta\mu_A} \over \sqrt{2}} + i {\sqrt{\eta\mu_B} \over \sqrt{2}} e^{i(\phi_{AB}-\pi/2)},\\
		\alpha_{D} &= i {\sqrt{\eta\mu_A} \over \sqrt{2}} + {\sqrt{\eta\mu_B} \over \sqrt{2}} e^{i(\phi_{AB}-\pi/2)},
	\end{aligned}
\end{equation}

\noindent where, again, the click probability is

\begin{equation}
	\begin{aligned}
		P_{\text{click}} = 1 - e^{-|\alpha|^2} (1-p_d),
	\end{aligned}
\end{equation}

\noindent from which we can again combine the click probabilities for C and D detectors and obtain the single-click probability

\begin{equation}
	\begin{aligned}
	P_{k_c,k_d}^Z(\mu_A,\mu_B,\phi_{AB})&=(P_{\text{click},C})^{k_c}(1-P_{\text{click},D})^{k_c}\\
	&\times (P_{\text{click},D})^{k_d}(1-P_{\text{click},C})^{k_d}
	\end{aligned}
\end{equation}

As mentioned above, we can divide the 2D domain of $\mu_A,\mu_B$ into arbitrary regions $S_{ij}$ playing the roles of ``decoy settings". The average gain in each region is:

\begin{equation}\label{eq:theory_gain}
	\begin{aligned}
	Q_{k_c,k_d,ij}^Z &= {1\over {2\pi P_{S_{ij}}}}\iint_{S_{ij}} \int_{0}^{2\pi} \\ &P_{k_c,k_d}^Z(\mu_A,\mu_B,\phi_{AB}) P_{int}(\mu_A,\mu_B)\\
	& d\mu_Ad\mu_B d\phi_{AB}.
	\end{aligned}
\end{equation}

\noindent where $P_{S_{ij}}$ is the normalization factor of intensity distribution:

\begin{equation}\label{eq:decoy_prob}
	\begin{aligned}
		P_{S_{ij}} &= \iint_{S_{ij}} P_{int}(\mu_A,\mu_B)d\mu_Ad\mu_B.
	\end{aligned}
\end{equation}

For instance, we can set a grid of $3\times 3$ square regions. In this case, the gain for each setting $(a,b)$ (where $a,b=1,2,3$) is

\begin{equation}
	\begin{aligned}
	Q_{k_c,k_d,ij}^Z &= {1\over {2\pi P_{S_{ij}}}}\int_{(a-1)\mu_{max}/3}^{a\mu_{max}/3} \int_{(b-1)\mu_{max}/3}^{b\mu_{max}/3} \int_{0}^{2\pi} \\ &P_{k_c,k_d}^Z(\mu_A,\mu_B,\phi_{AB}) P_{int}(\mu_A,\mu_B)\\
	& d\mu_Ad\mu_B d\phi_{AB}.
	\end{aligned}
\end{equation}

An example decoy setting strategy is shown in main text Fig. 2 (b).

\subsection{Characterization of Source States}

By the way, just for reference, we can also characterize the states that are sent out by Alice and Bob. The information below is not used in the simulation (which only uses integration of the observable functions and do not directly use the density matrix), though, and it is listed only to aid the readers' understanding of the setup.

Before they perform post-selection, the state Alice and Bob each send is simply a globally-phase-randomized and intensity-randomized coherent state, similar to the output of a passive-decoy setup. The intensity probability distribution of Alice's (or Bob's) output is $P_{int}(\mu)=1/(\pi \sqrt{\mu(\mu_{max}-\mu)})$ \cite{FullyPassiveThis}. The joint state, when expressed in the Fock basis, can be written as:

\begin{equation}\label{eq:probability}
	\begin{aligned}
		\rho_{AB} &= \sum_{n_A,n_B} (\int_{0}^{\mu_{max}} \int_{0}^{\mu_{max}} P_{int}^{2D}(\mu_A,\mu_B) \\
		&\times P_{Poisson}(\mu_A,n_A)P_{Poisson}(\mu_B,n_B)d\mu_A d\mu_B )\\
		&\times \ket{n_A,n_B}\bra{n_A,n_B}
	\end{aligned}
\end{equation}

\noindent where $P_{int}^{2D}$ is defined in Eq. \ref{eq:pint} and $P_{Poisson}$ is the Poissonian distribution $P_{Poisson}(\mu,n)=exp(-\mu)\mu^{n}/n!$.

In the Z basis, Alice and Bob each post-select a range of intensities as their decoy setting (while keeping the global phase random). The conditional states can be written as:

\begin{equation}
	\begin{aligned}
		\rho_{AB}^{Z} &= \sum_{n_A,n_B} ({1\over {P_{S_{ij}}}}\int_{(a-1)\mu_{max}/3}^{a\mu_{max}/3} \int_{(b-1)\mu_{max}/3}^{b\mu_{max}/3} \\ 
		&\times P_{int}(\mu_A,\mu_B)P_{Poisson}(\mu_A,n_A)P_{Poisson}(\mu_B,n_B) \\
		&\times d\mu_A d\mu_B )\\
		&\times \ket{n_A,n_B}\bra{n_A,n_B}
	\end{aligned}
\end{equation}

\noindent where $S_{ij}$ is the decoy region corresponding to setting $(a,b)$ (where, as an example, $a,b=1,2,3$ and square regions are used).\\

In the X basis, Alice and Bob post-process the data and divide them into pairs of slices. Here for simplicity, let us focus on Alice, and consider the case where she selects the first slices for her sources $A1$ and $A2$, i.e. $k_{A1}=k_{A2}=1$. The case where she selects differently indexed slices and the case for Bob can both be derived from this basic case straightforwardly.

Based on Eq. \ref{eq:source_interference}, we know that if Alice inputs two states $\ket{\alpha_0 e^{i\phi_{A1}}}$ and $\ket{\alpha_0 e^{i\phi_{A2}}}$ into a beam splitter, on the output port A3, the coherent state can be described by the complex amplitude $\alpha_{A3} = {\alpha_0 \over \sqrt{2}} e^{i\phi_{A1}} + i {\alpha_0 \over \sqrt{2}} e^{i(\phi_{A2}-\pi/2)}$. Let us denote it as $\alpha_{A3}(\phi_{A1},\phi_{A2})$ as it is a function of the input phases $\phi_{A1}$ and $\phi_{A2}$. The output state of Alice (conditional to her choosing $k_{A1}=k_{A2}=1$ in the X basis) is:

\begin{equation}
	\begin{aligned}
		\rho_{A}^{X} &= {1\over {4\Delta_{\phi}^2}}\int_{-\Delta_{\phi}}^{\Delta_{\phi}}\int_{-\Delta_{\phi}}^{\Delta_{\phi}} \ket{\alpha_{A3(\phi_{A1},\phi_{A2})}}\bra{\alpha_{A3(\phi_{A1},\phi_{A2})}} \\
		&\times d\phi_{A1}d\phi_{A2}\\
		&= \sum_{n_A,n'_A} ({1\over {4\Delta_{\phi}^2}}\int_{-\Delta_{\phi}}^{\Delta_{\phi}}\int_{-\Delta_{\phi}}^{\Delta_{\phi}} e^{-|\alpha_{A3}(\phi_{A1},\phi_{A2})|^2}\\
		&\times {{\alpha_{A3}(\phi_{A1},\phi_{A2})^{n_A} \alpha_{A3}^*(\phi_{A1},\phi_{A2})^{n'_A}}\over \sqrt{n_A!n'_A!}}d\phi_{A1}d\phi_{A2})\\
		&\times \ket{n_A}\bra{n'_A}
	\end{aligned}
\end{equation}

%%%%%%%%%%%%%%%%%
%%%%%%%%%%%%%%%%%
%%%%%%%%%%%%%%%%%

From the above, we can similarly derive $\rho_B^X$ and hence $\rho_{AB}^X=\rho_A^X \otimes \rho_B^X$.

Note that, here $\rho_A^X$ is the same state being sent out by both the physical setup in main text Fig. 3 and the equivalent setup with a virtual ``local channel" in main text Fig .4, since by definition both setups interfere two coherent states $\ket{\alpha_0 e^{i\phi_{A1}}}$ and $\ket{\alpha_0 e^{i\phi_{A2}}}$ at a beam splitter, where $\phi_{A1}$ and $\phi_{A2}$ each fluctuate within a range of $[-\Delta_{\phi},\Delta_{\phi}]$. The only difference is that, in the former setup, the fluctuations come from the random phase initialization in the two sources (after which the phases are further post-selected into two given slices), while in the latter setup, the fluctuations come from the active modulation of two identical incoming pulses being split off from the same source.

In the case of active TF-QKD, the X basis signal states are simply coherent states $\ket{\alpha}$ and $\ket{-\alpha}$. On the other hand, in the case of passive TF-QKD, the fluctuations of the phases result in the state $\rho_A^X$ with mixed phases (within the range $[\Delta_{\phi},\Delta_{\phi}]$) and mixed intensities (possible to take values smaller than $\mu_{max}$, when $\phi_{A1}$ and $\phi_{A2}$ are mismatched). 

\begin{figure}[h]
	\includegraphics[scale=0.5]{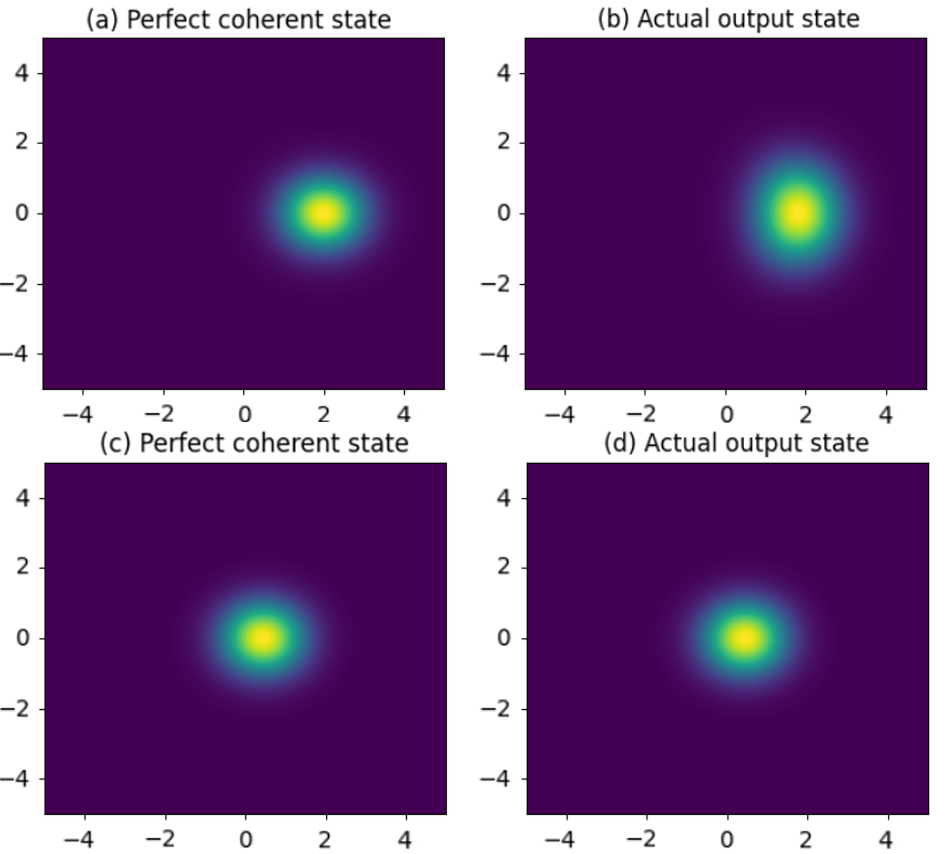}
	\caption{(a) The Wigner function of a coherent state with intensity $\mu=2$. (b) The Wigner function of $\rho_A^X$ with intensity $\mu_{max}=2,\Delta_{\phi}=\pi/4$. (c) The Wigner function of a coherent state with intensity $\mu=0.1$. (d) The Wigner function of $\rho_A^X$ with intensity $\mu_{max}=0.1,\Delta_{\phi}=\pi/24$ (corresponding to $N=24$). We deliberately choose a large intensity and wide phase slice for the above two figures to illustrate the spread of phases for $\rho_A^X$ due to the finite size of the phase slice. For the below two figures with reasonable intensity and phase slice values, we can see that the state $\rho_A^X$ is very close to the perfect coherent state (and the Wigner functions are almost indistinguishable). The Wigner function plots are generated with the QuTiP library \cite{qutip}.}
	\label{fig:a1b}
\end{figure} 

To visualize the state, we can plot out the Wigner function of $\rho_A^X$, as shown in Fig. \ref{fig:a1b}, where we also plot the Wigner function of a coherent state with intensity fixed at $\mu_{max}$ for comparison (here $\mu_{max}=2|\alpha_0|^2$). In Fig. \ref{fig:a1b} (a)(b) we choose an extreme case of $\mu_{max}=2,\Delta_{\phi}=\pi/4$ to more clearly illustrate the mixed phases caused by post-selection into phase slices of a finite size.

In Fig. \ref{fig:a1b} (c)(d) we choose a set of reasonable values $\mu_{max}=0.1,\Delta_{\phi}=\pi/24$ (which are similar to the parameters used in our simulations). As can be seen, the output signal state in the passive TF-QKD setup is actually quite similar to that of active TF-QKD. Here when $N=24$, the fidelity between $\rho_A^X$ and $\ket{\alpha}\bra{\alpha}$ is as high as approximately $99.97\%$, and even when $N=8$ the fidelity is still about $99.74\%$. This means that the finite size of phase slices has little effect on the actual statistics measured by Alice and Bob (hence the QBER is not affected much), but rather is mainly a security concern, which we address by constructing the virtual local channels and revising the photon-number yields (hence also the phase error rate), as described in the previous section.

	\section{Choice of Number of Phase Slices}

\begin{figure}[h]
	\includegraphics[scale=0.225]{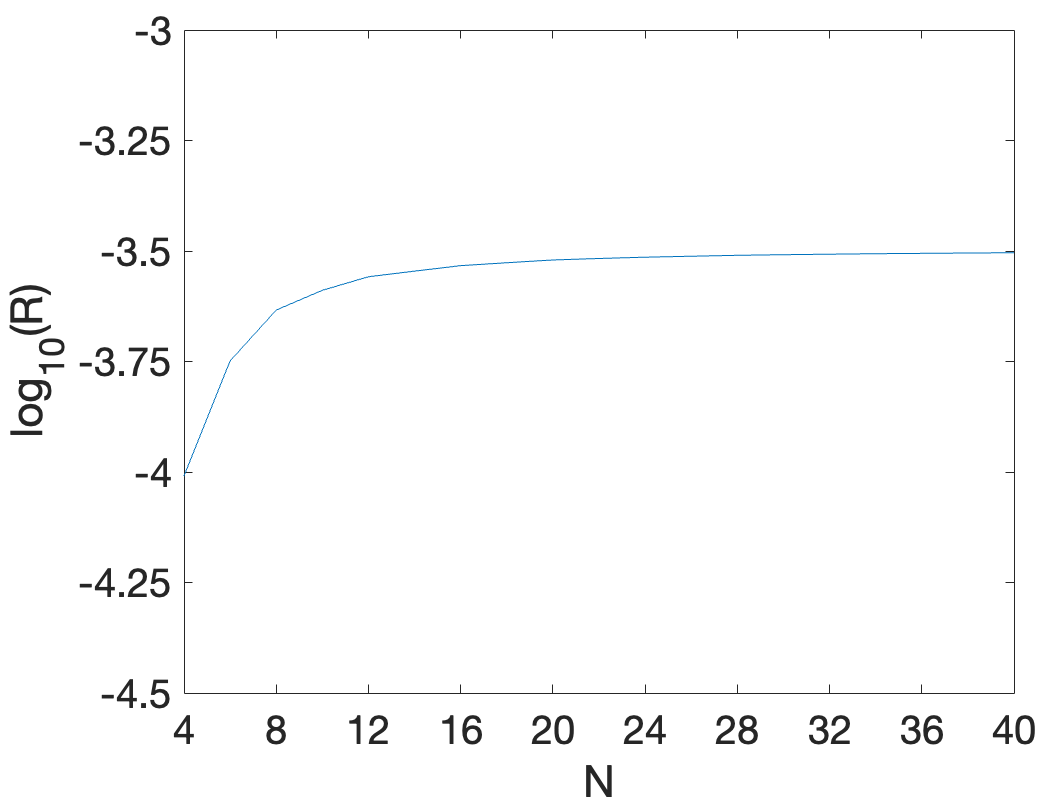}
	\caption{Comparison of key rate using different values of phase slice number N for the infinite data, infinite decoys scenario. Distance $L_{AB}$ is fixed at 100km and signal intensity is fixed at $\mu_{max}=0.018$ (same as the optimal $\mu_{max}$ used in main text Fig. 5 at 100km for $N=24$). As can be seen, the key rate always increases with larger N, but has diminishing returns when N becomes too large. This means that choosing a reasonably large value for N will already yield a sufficiently high key rate, e.g. at 100km, using $N=24$ the key rate is $3.06\times 10^{-4}$ while using $N=40$ the key rate is $3.13\times 10^{-4}$, merely $2.3\%$ higher.}
	\label{fig:a5}
\end{figure}

In this section we briefly discuss how the number of phase slices N is chosen. As mentioned in the main text, in the asymptotic scenario, the key rate never decreases when we increase N, since we add up all phase slice combinations, and further subdividing each slice only yields more information but no less sifting. In principle, this means that N is only limited by the resolution and accuracy of our local phase detection devices and we can set as large an N as our devices allow. Nonetheless, as shown in Fig. \ref{fig:a5}, when N is sufficiently large (say $N\geq 20$), further increasing the slice number has diminishing return, yet it will increase the computation cost since the key rate is calculated for $N^3$ times. In the simulation in main text Fig. 5, we choose a reasonably large value of $N=24$.

\section{Frequency Drift Compensation}

Stabilizing the frequencies between independent laser sources is a challenge for TF-QKD in general (both for active and passive systems) due to the sensitivity of the protocol to phase shift, which directly contributes to QBER. This is particularly a challenge for passive TF-QKD, since we require independence and randomness for phases in each of Alice's and Bob's sources, which prevent the use of phase locking, a commonly used technique in active TF-QKD. In this section we propose possible techniques to address the frequency drift problem for passive TF-QKD.

\begin{figure}[h]
	\includegraphics[scale=0.3]{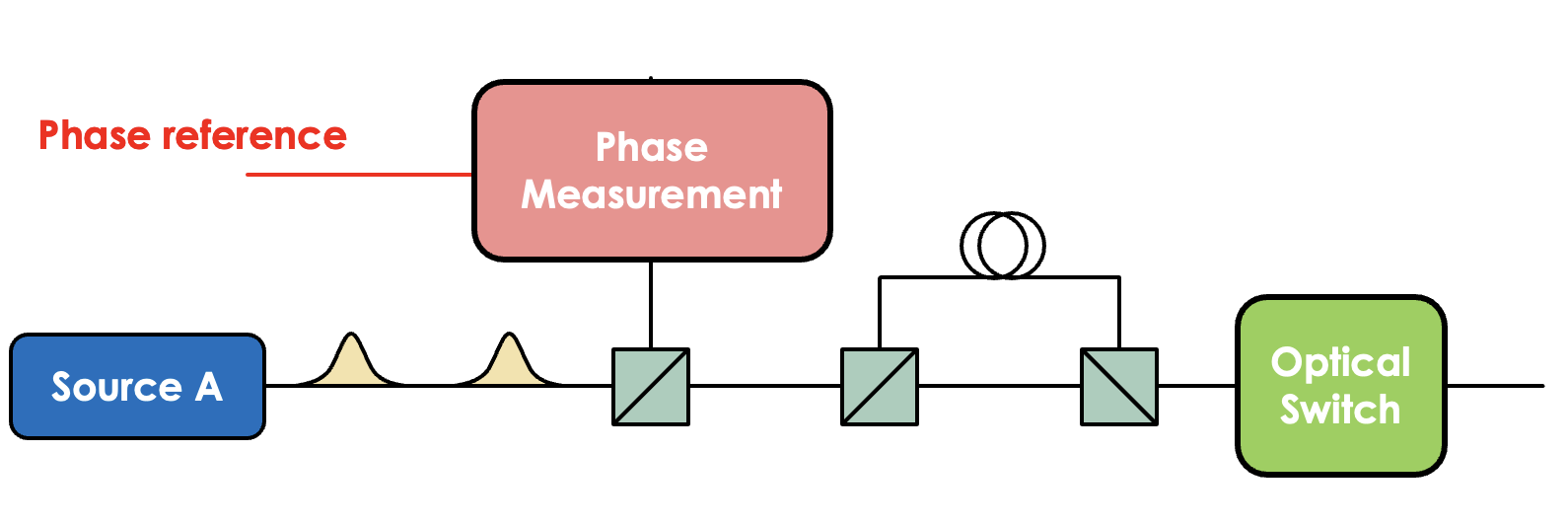}
	\caption{Single laser scheme for local frequency stability between pulses (for Alice as an example), similar to the scheme used in Ref. \cite{FullyPassiveThis} for fully passive BB84 (such a technique was first proposed in the original passive decoy-state QKD paper \cite{passivedecoy2}). A single laser source combined with delay lines allows neighboring pulses (which have independent and random phases) to interfere, simulating the effect of two independent sources A1 and A2. Every other pulse at the output port needs to be suppressed by an optical switch or intensity modulator so that the final output pulses have no phase correlation. The phase measurement can be performed before the interference so that Alice knows the exact phase of each pulse.}
	\label{fig:a3}
\end{figure}

\begin{figure*}[t]
	\includegraphics[scale=0.38]{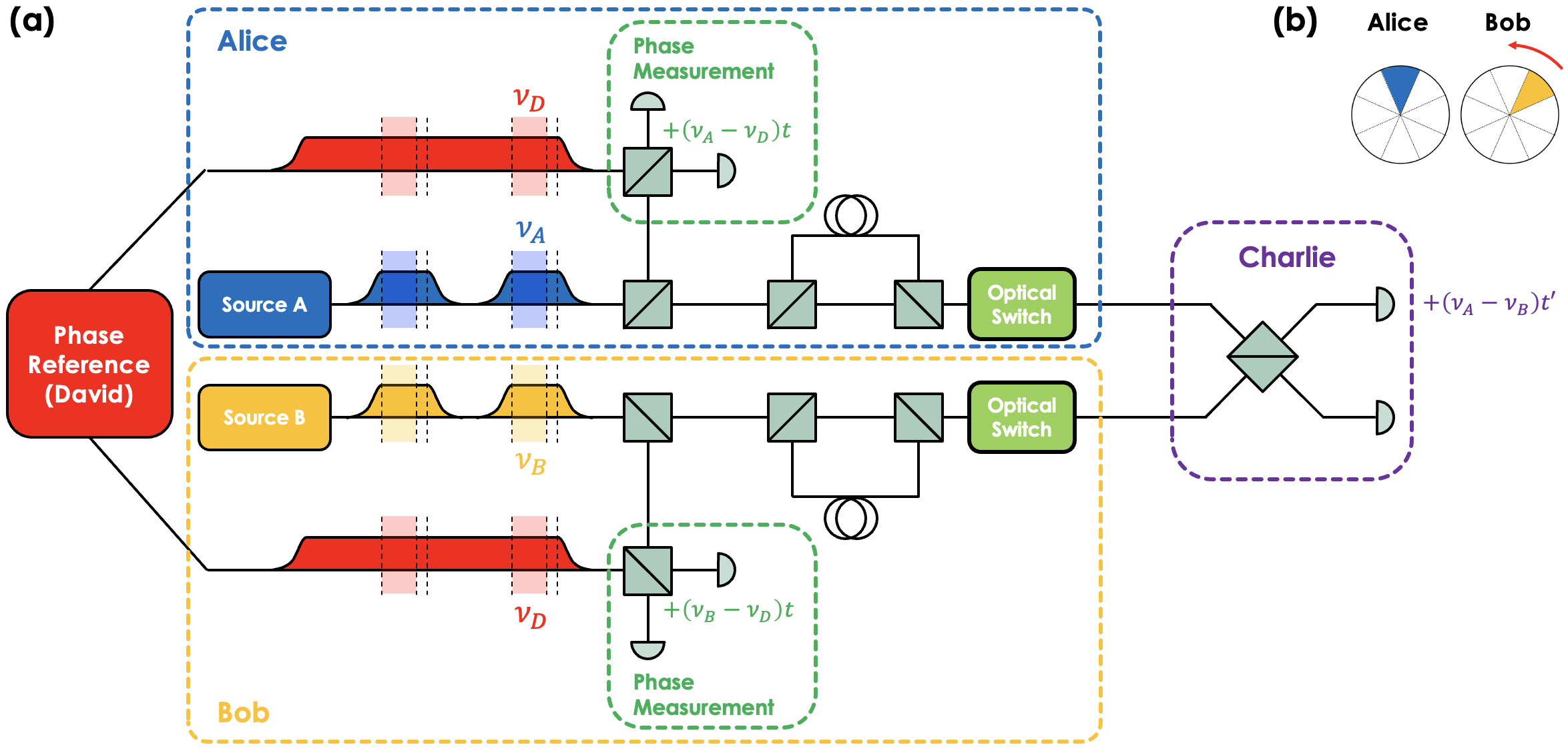}
	\caption{(a) Setup of the frequency compensation scheme for remote parties in passive TF-QKD based on post-processing, inspired by the technique proposed in Ref. \cite{nophaselocking2}. The station David sends reference lasers to Alice and Bob (calibrated such that Alice and Bob share a common phase reference and the same frequency $\nu_D$ for their reference signals). Alice and Bob use gain-switched lasers (which we assume to have frequencies $\nu_A,\nu_B$ during a given pulse window) and respectively set aside part of each pulse for local calibration, where they track the intensity fluctuation of their local detectors and record the beat frequencies $\Delta_{AD}=\nu_A-\nu_D$ and $\Delta_{BD}=\nu_B-\nu_D$. Then, they can announce their recorded frequencies and derive $\Delta_{AB}=\Delta_{AD}-\Delta_{BD}=\nu_A-\nu_B$. Combining it with the travel time to Charlie $t'$, they can predict the phase drift $\Delta_{AB} t'$ caused by Alice and Bob's frequency mismatch. (b) After Alice and Bob predict the phase drift at Charlie due to their frequency mismatch, Bob can simply rotate his frame of reference by the same amount during post-processing to compensate for the drift. The compensation is performed per-pulse (i.e. each pulse can in principle have a different drift amount, which can be tracked by Alice and Bob's local calibration process).}
	\label{fig:a4}
\end{figure*} 

Firstly, we remark that local frequency drift between sources A1 and A2 (B1 and B2) is relatively easy to address if we apply a technique used in Ref. \cite{FullyPassiveThis} for fully passive BB84 and originally in Ref. \cite{passivedecoy2} for passive decoy states: A single laser source combined with adequate delay lines allows neighboring pulses to interfere, simulating the effect of independent lasers. An active switch or intensity modulator needs to be used to suppress even/odd pulses, but there is no additional side-channel introduced since the modulation follows a fixed pattern. Also, as explained in Ref. \cite{FullyPassiveThis}, the security can be quantified even when taking into consideration the finite extinction ratio of the switch/intensity modulator, by treating insufficiently suppressed pulses as a Trojan Horse Attack, for which it is shown that 60-80dB of extinction ratio (e.g. from concatenated intensity modulators) will limit the effect on key rate to a minimum. An illustration of the single laser scheme can be found in Fig. \ref{fig:a3}.

Secondly, we propose one possible scheme to address the remote frequency drift between Alice and Bob. This scheme is inspired by the technique proposed in Ref. \cite{nophaselocking2}, which implements TF-QKD without phase locking and compensates for phase drift caused by frequency drift using classical post-processing only. An illustration of our scheme for remote frequency drift compensation in passive TF-QKD is shown in Fig. \ref{fig:a4} (a). Here we can assume Alice and Bob each uses a single laser (combined with a delay line to simulate two local sources). A classical reference laser (sent from the station David, which could also be Charlie's station) is distributed to each of Alice and Bob. Such a reference laser is utilized in many active TF-QKD experiments too for common phase reference and there should be no additional technical challenge that is particular to passive TF-QKD here. Therefore, we can assume that, after the calibration at David, the reference signals that Alice and Bob receive have the same frequency and phase reference point. Starting from here, the actual steps of the compensation are listed below (note that, while the local calibration is performed during the communication process, all of the announcement and compensation mentioned below are performed during post-processing):

1. Alice (Bob) generates phase randomized pulses with a local gain-switched laser, whose frequency at a given pulse window can be denoted as $\nu_A$ ($\nu_B$). Alice (Bob) sets aside part of the duration of each pulse for classical local calibration and leaves the remaining part as the signal. The calibration part can be suppressed when outputting. In the phase measurement module, she (he) interferes her (his) signal with the incoming reference signal from David (which has frequency $\nu_D$ during the pulse window). During the calibration, due to the frequency difference $\Delta_{AD}=\nu_A-\nu_D$ ($\Delta_{BD}=\nu_B-\nu_D$), the measured intensity would fluctuate with the beat frequency, e.g. $I_{A0}={1\over 2}I_{max}[1+cos(\phi_{AD}+\Delta_{AD} t)]$, where $I_{A0}$ is the intensity at one of Alice's local classical detectors, $\phi_{AD}$ is the phase difference between Alice and David and $t$ is the elapsed time during the calibration period.

2. Alice and Bob can announce and compare their beat frequencies within the same pulse window, $\Delta_{AD}$ and $\Delta_{BD}$, and derive $\Delta_{AB}=\Delta_{AD}-\Delta_{BD}=\nu_A-\nu_B$, which is the frequency difference between Alice and Bob's local lasers. With this information (and calculating the time $t'$ to travel from Alice/Bob to Charlie), they can derive the phase shift $\Delta_{AB} t'$ at Charlie due to the frequency difference between Alice and Bob. There could also be phase shift due to channel noise, but such shift can be compensated in the same way as in active TF-QKD (e.g. via feedback) and is not a unique problem to passive TF-QKD.

3. After Alice and Bob derive the phase shift $\Delta_{AB} t'$, Bob can simply rotate his frame of reference (i.e. the zero point on the circumference of his phase slices) by the same amount to match the phase drift, as shown in Fig. \ref{fig:a4} (b). The compensation process can be performed entirely during post-processing. \footnote{In a way, this is again similar to the compensation process of PM-QKD, which also allows Bob to freely rotate his reference point, making the protocol immune to any \textit{fixed} misalignment. As mentioned in the main text, passive TF-QKD also has such immunity to fixed phase misalignment, since it naturally prepares signals in all directions between $[0,2\pi)$ and adds up the key rate from all slice combinations. However, neither PM-QKD nor passive TF-QKD by default allows for phase differences \textit{changing} with time. Therefore, the purpose of the frequency drift compensation scheme here is to cancel out the real-time phase drift \textit{per pulse}, after which any fixed misalignment is not a problem for passive TF-QKD.}

One main difference between our proposal here and that of Ref. \cite{nophaselocking2} is that the latter performs the frequency tracking by combining single-photon detector clicks with fast fourier transform (FFT) to guess the frequency (since the tracking is performed at Charlie), so the process should actually be considerably easier for our scenario since we are using local detectors working at classical intensity levels (which output continuous voltage values) at Alice's and Bob's local stations.

Another difference between the schemes is that Alice and Bob are using gain-switched lasers (while David can still use a continuous-wave reference laser for simplicity) in the passive scheme. This means that we need to perform per-pulse calculation of frequency difference, rather than calculating every few thousands of pulses (every few microseconds) as implemented in Ref. \cite{nophaselocking2}, since each pulse can have slightly different frequencies. Considering the frequency difference (i.e. beat frequency) is on the order of 100MHz as mentioned in Ref. \cite{nophaselocking2}, as long as we keep the pulse lengths to a reasonable number (say 100ns), we should be able to have multiple fluctuation periods to estimate the beat frequency. Also, considering Ref. \cite{nophaselocking2} performed calibration and assumed constant frequencies across microseconds for continuous-wave lasers, if we use gain-switched lasers and truncate/post-select the pulses to only measure stable regions (to avoid chirping at the beginning of each pulse), we should also have a stable frequency at least throughout the pulse (which is on the order of only say 100 ns), such that we can assume that the frequency differences remain constant values for each pulse window in the estimation.

{\color{black}
\section{Finite-Size Analysis}

In this section we outline a preliminary finite-size analysis for our new passive TF-QKD protocol. Note that this analysis is only meant to estimate the performance of the new protocol with finite data size in order to show its practicality, while a more rigorous proof will be the subject of future studies.

Here we will follow the analysis from Ref. \cite{finiteTF1} for the CAL TF-QKD protocol, which uses Chernoff's bound to upper-bound the photon number yields in decoy-state analysis and uses a concentration inequality (similar to but tighter than Azuma's inequality) to upper-bound the phase error rate based on photon number yields. Note that, in principle, other finite-size analysis for TF-QKD can be applied too, such as Ref. \cite{finiteTF2} which is also for CAL TF-QKD (although the proof may need modifications since the vacuum decoy state for passive TF-QKD is always non-zero) or NPP TF-QKD \cite{finiteTF3}, which uses the same physical states as CAL TF-QKD.

Our passive protocol has two main differences from its active counterpart: 

\begin{enumerate}
	\item The intensities of decoy states are prepared in continuous intervals rather than discrete values. This simply means a photon number state is, instead of coming from a single random draw from the Poissonian distribution, now coming from two consecutive random draws where one first picks an intensity from the intensity distribution and then picks a photon number from the corresponding Poissonian distribution. The process of Alice/Bob choosing a certain photon number state to send (and the probability of doing so) is still independent in each round, regardless of Eve's attacks. Therefore the same Chernoff's bound can be applied to bound the yield of each photon number state $Y_{mn}$.
	\item The signal states have mixed phases and intensities. However, as discussed in Appendix D, we assume that \textit{perfect} signal states $\ket{\alpha},\ket{-\alpha}$ are prepared and are sent through local channels that are attributed to Eve (thus the local channels are considered as environmental noise/loss). Therefore, the security analysis for our protocol (e.g. the bounding of phase error rate in the signal X basis based on Z basis statistics) should be no different from that of the active case.
\end{enumerate}

Therefore, fortunately, we can reuse the majority of the analysis as-is from Ref. \cite{finiteTF1}. However, there is one main incompatibility, namely that Ref. \cite{finiteTF1} requires the signal intensity $|\alpha|^2$ to be smaller than the largest decoy intensity:

\begin{equation}
	|\alpha|^2 \leq max\{\mu_i\}
\end{equation}

\noindent where $\{\mu_i\}$ is the set of decoy intensities used. This is a specific requirement (not present in the asymptotic case) that is imposed by the analysis of Ref. \cite{finiteTF1} to ensure that the residue phase error contributions from the infinitely many photon number states outside the photon number cut-off region $S_{cut}$ still converge. Such a requirement, however, is a problem for passive TF-QKD, where both the signal states and decoy states are prepared from the same source setup, and they are only differentiated by post-selection, resulting in identical maximum intensities for the signal/decoy states, i.e. $\mu_{max}$. This makes our asymptotic passive TF-QKD protocol introduced so far incompatible with a naive application of the analysis in Ref. \cite{finiteTF1} since the phase error would diverge.

To overcome this incompatibility, below we will present (1) an improved local channel model for Alice's/Bob's sources that enables smaller signal intensities in the virtual ``perfect" source (instead of always fixing it to $\mu_{max}$), as well as (2) a revised post-selection strategy to further increase Alice's/Bob's ability to control the intensities of their post-selected signals. These two modifications allow us to apply the analysis in Ref. \cite{finiteTF1} to calculate the key rate of passive TF-QKD under finite-size effects.

\subsection{Improved Local Channel Model}

In this subsection, we propose an improved local channel model for passive TF-QKD. Compared with the unbalanced Mach-Zehnder Interferometer (MZI) model with two phase modulations in the main text, the new model replaces them with a single beamsplitter and a global phase modulation. This allows us to explicitly track the minimum and maximum transmittances of the local beamsplitter (i.e. track the output signal intensity) as well as to perform different post-selection strategies, such as post-selecting based on the transmittance, instead of the two local phase modulations. Note that, this is entirely an improvement in the theoretical model, while the physical setup is identical as we described in the main text.

As a recapitulation, in the analysis for asymptotic passive TF-QKD described in this paper so far, for the signal X basis we have always assumed virtual \textit{perfect sources} of a fixed intensity $\mu_{max}$, and any phase and intensity fluctuations caused by finite post-selection regions are considered as the effect of a local channel inside Alice's/Bob's lab. The source model is shown in Fig. \ref{fig:a7} (a), which is a reproduction of the same plot as Fig. 3 in the main text \footnote{\color{black}Except that, for simplicity, in the main text we showed an example where Alice's phase slice indices $k_{A1}$ and $k_{A2}$ happened to be exactly zero (i.e. average phase is zero), while here we describe the general case of any position of the two phase slices} \footnote{Another subtlety worth mentioning here is that, in the main text Fig. 1 (a) and Fig. 3, to keep the figures simple, we have implicitly incorporated two constant $-\pi/2$ phase shifts into $\phi_{A1}$ and $\phi_{A2}$, while here in Fig. \ref{fig:a7} (a) we explicitly draw out the phase shifts to derive the precise expressions for the channel model. The phase shifts cancel out the $\pi/2$ phase shifts in the reflected beams at the first and the second beamsplitter, such that the output port reaches maximum intensity when $\phi_{A1}=\phi_{A2}$. Note that either including or omitting these two phase shifts above does not change any numerical results, since $\phi_{A1},\phi_{A2}$ are uniformly distributed between $[0,2\pi)$, so any constant shift simply means changing the definition of the reference zero-point on the circumference.}.

\begin{figure}[t]
	\includegraphics[scale=0.315]{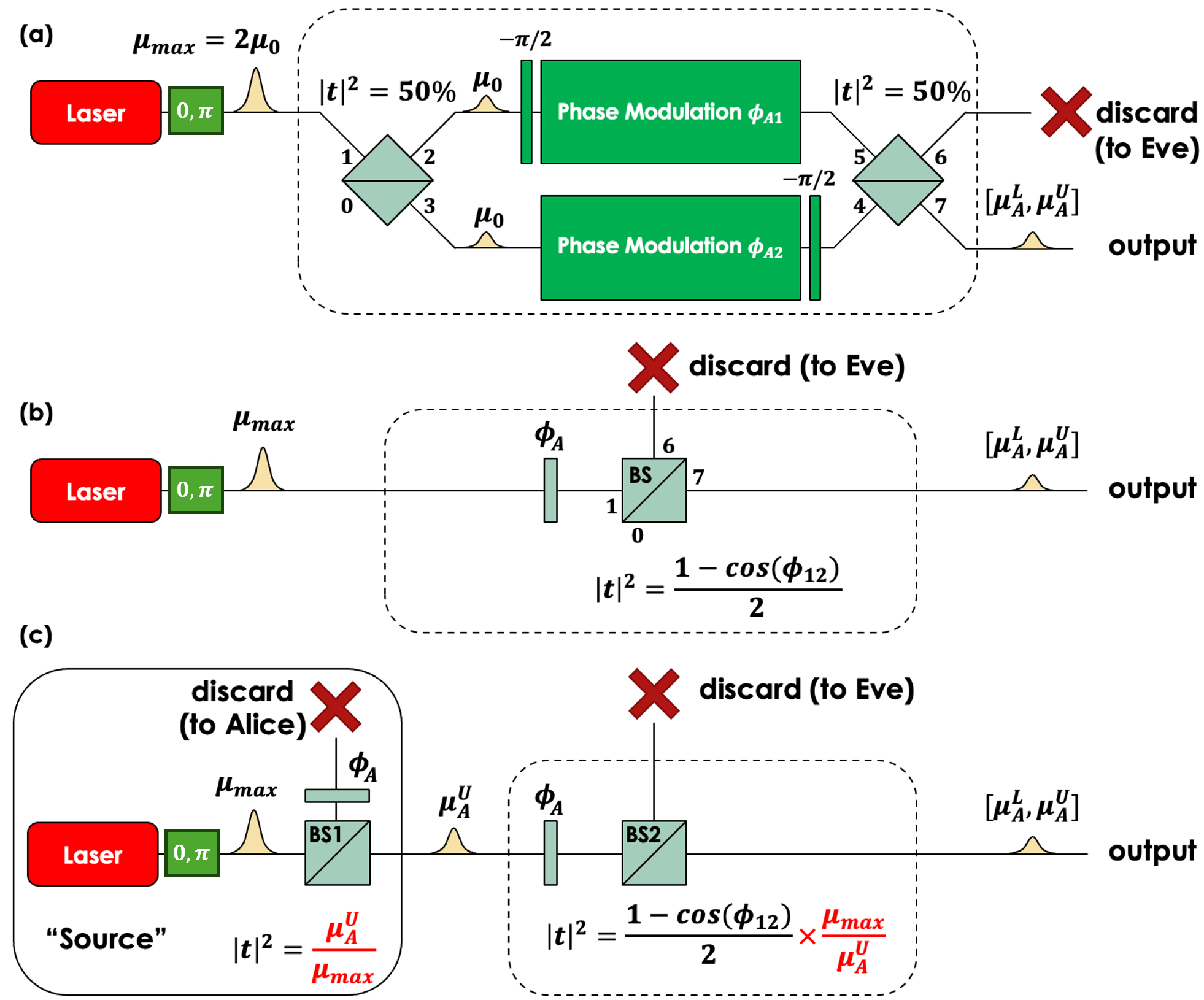}
	\caption{{\color{black}(a) Original local channel model in the main text, based on an unbalanced Mach-Zehnder Interferometer with two variable phase modulations, $\phi_{A1}$ and $\phi_{A2}$. The phase modulations affect the output intensity $\mu_A$, which, say, lies within some region $[\mu_A^L,\mu_A^U]$, depending on Alice's post-selection strategy on $\phi_{A1},\phi_{A2}$ (say, selecting phase slices with indices $k_{A1},k_{A2}$). Of course, the output never exceeds the input, i.e. $\mu_A^U\leq \mu_{max}$. (b) An equivalent local channel model based on a single variable beamsplitter and a global phase shifter, where the transmittance lies within $[\mu_A^L/\mu_{max},\mu_A^U/\mu_{max}]$. (c) Another local channel model, where the local loss is split into a constant part $\mu_A^U/\mu_{max}$ and a variable part that lies within $[\mu_A^L/\mu_A^U,1]$. The constant part of the loss can be incorporated into the trusted local source, i.e. assuming a weaker signal intensity $\mu_A^U$ to begin with. This model still outputs the same physical signals as the above two but has a less pessimistic security boundary assumption (thus potentially providing better key rate) as it only attributes part of the local noise to Eve.}}
	\label{fig:a7}
\end{figure} 

Here we focus on one user (say Alice), as the treatment for the other user's source is identical. In our model so far, each single event (Alice's two lasers emitting two pulses with random phases $\phi_{A1}$ and $\phi_{A2}$) can be equivalently considered as a signal from a single virtual source with intensity $\mu_{max}=2\mu_0$ going through a 50-50 beamsplitter and undergoing two phase modulations of $\phi_{A1}$ and $\phi_{A2}$ on the two arms, before being recombined at another 50-50 beamsplitter. Alice can observe $\phi_{A1},\phi_{A2}$ and perform post-selection upon them (e.g. defining phase slices). For each given $\phi_{A1},\phi_{A2}$, the conversion between the input and output modes can be written as

\begin{comment}
\begin{equation}
	\begin{aligned}
		a^{\dagger}_1 &\rightarrow {1\over \sqrt{2}} (ia_2^\dagger + a_3^\dagger) \\
		&\rightarrow {1\over \sqrt{2}} \left[e^{i(\pi/2 + \phi_{A1})}a_5^\dagger + e^{i\phi_{A2}} a_4^\dagger\right] \\
		&\rightarrow {1\over 2} \left[e^{i(\phi_{A1}+\pi)}+e^{i\phi_{A2}}\right]a_6^\dagger \\
		&+ {1\over 2} \left[e^{i(\phi_{A1}+\pi/2)}+e^{i(\phi_{A2}+\pi/2)}\right]a_7^\dagger\\
	\end{aligned}
\end{equation}
\end{comment}

\begin{equation}\label{eq:BS}
	\begin{aligned}
		a^{\dagger}_1 &\rightarrow {1\over 2} \left[e^{i(\phi_{A1}+\pi/2)}+e^{i(\phi_{A2}-\pi/2)}\right]a_6^\dagger \\
		&+ {1\over 2} \left[e^{i\phi_{A1}}+e^{i\phi_{A2}}\right]a_7^\dagger\\
		a^{\dagger}_0 &\rightarrow {1\over 2} \left[e^{i\phi_{A1}}+e^{i\phi_{A2}}\right]a_6^\dagger \\
		&+ {1\over 2} \left[e^{i(\phi_{A1}-\pi/2)}+e^{i(\phi_{A2}+\pi/2)}\right]a_7^\dagger\\
	\end{aligned}
\end{equation}

\noindent For a coherent state \cite{passivedecoy1}, 

\begin{equation}
	\begin{aligned}
		\ket{0}_0 \ket{\sqrt{\mu_{max}}}_1 \rightarrow &\left| {\sqrt{\mu_{max}}\over 2} \left[e^{i(\phi_{A1}+\pi/2)}+e^{i(\phi_{A2}-\pi/2)}\right] \right \rangle_6 \\
		&\left| {\sqrt{\mu_{max}}\over 2} \left(e^{i\phi_{A1}}+e^{i\phi_{A2}}\right) \right \rangle_7 \\
	\end{aligned}
\end{equation}

\noindent The intensity and phase of the output signal satisfy

\begin{equation}\label{eq:MZI}
	\begin{aligned}
		\mu_A&={{1+\cos({\phi_{A2}-\phi_{A1}})}\over 2}\mu_{max}={{1+\cos{\phi_{21}}}\over 2}\mu_{max}, \\
		\phi_A&={{\phi_{A2}+\phi_{A1}}\over 2},\\
	\end{aligned}
\end{equation}
\noindent where $\phi_{21}=\phi_{A2}-\phi_{A1}$.

The big problem so far is that, based on the above model, no matter how Alice post-selects upon her $\phi_{A1}$ and $\phi_{A2}$, her source intensity is always assumed to be a constant $\mu_{max}$, since all modulations are considered effects of the channel, not the source. For instance, in the extreme case, suppose that Alice uses very narrow phase slice sizes such that the output signal is almost constant with little fluctuation, and she chooses two phase slices whose center positions $\phi_{A1}^{avg},\phi_{A2}^{avg}$ are approximately $\pi/2$ apart, the final output intensity would then be roughly $\mu_{max}/2$ all the time. However, we will still have to assume the source intensity is $\mu_{max}$, and the reduction in intensity (about 3dB of nearly-fixed loss) is attributed to the local channel and thus to Eve. This is obviously a too pessimistic assumption.

To address this problem and enable potentially smaller signal intensities in our analysis, we first simplify the local channel model and decouple the local loss and phase shifts. We observe that the MZI setup in Fig. \ref{fig:a7} (a) is equivalent to a single unbalanced beamsplitter with

\begin{equation}
	\begin{aligned}
		t_0 &= {1\over 2} \left[e^{i\phi_{A1}}+e^{i\phi_{A2}}\right]\\
		r_0 &= {1\over 2} \left[e^{i(\phi_{A1}-\pi/2)}+e^{i(\phi_{A2}+\pi/2)}\right]\\
		t_1 &= {1\over 2} \left[e^{i\phi_{A1}}+e^{i\phi_{A2}}\right]\\
		r_1 &= {1\over 2} \left[e^{i(\phi_{A1}+\pi/2)}+e^{i(\phi_{A2}-\pi/2)}\right]\\
	\end{aligned}
\end{equation}

\noindent which provides the same conversion relations between the input and output modes as in Eq. \ref{eq:BS}:

\begin{equation}
	\begin{aligned}
		&a_1^\dagger \rightarrow r_1 a_6^\dagger + t_1 a_7^\dagger \\
		&a_0^\dagger \rightarrow t_0 a_6^\dagger + r_0 a_7^\dagger \\
	\end{aligned}
\end{equation}

\noindent We can check that $|r_0|^2 + |t_0|^2 = 1$ and $|r_1|^2 + |t_1|^2 = 1$, as well as $r_0^*t_1 + r_1 t_0^* = 0$ and $r_0^*t_0 + r_1 t_1^* = 0$. Also, we have the freedom to multiply a global phase factor of $e^{-i\phi_A}$ simultaneously to $t_0,r_0,t_1,r_1$ (which maintains the beamsplitter conditions) and move the constant phase shift of $\phi_A$ in front of the input ports. If we set $\phi_G = (\phi_{A1}+\phi_{A2})/ 2$, then the beamsplitter has real coefficients 

\begin{equation}
	\begin{aligned}
		t_0 &= \cos[(\phi_{A2}-\phi_{A1})/2]\\
		r_0 &= - \sin[(\phi_{A2}-\phi_{A1})/2]\\
		t_1 &= \cos[(\phi_{A2}-\phi_{A1})/2] \\
		r_1 &= \sin[(\phi_{A2}-\phi_{A1})/2] \\
	\end{aligned}
\end{equation}

\noindent which preserves the input-output phase relations between coherent states.

We can see that, at this point, the local channel has been decoupled into a single ``pure-loss" beamsplitter whose transmittance is $|t_1|^2={{[1+\cos(\phi_{A2}-\phi_{A1})]}/ 2}$ as well as a global phase shift $\phi_A=(\phi_{A1}+\phi_{A2})/ 2$. This is consistent with the output state of the MZI described in Eq. \ref{eq:MZI}. Such an equivalence between an unbalanced MZI and a single unbalanced beamsplitter has been observed in previous literature, such as \cite{fingerprinting}. The single beamsplitter setup is shown in Fig. \ref{fig:a7} (b). Depending on the values of $\phi_{A1}$ and $\phi_{A2}$ (and the post-selection regions), the output intensity out of Alice's lab lies within a range $[\mu_A^L,\mu_A^U]$, in other words the transmittance of the local beamsplitter lies within $[\mu_A^L/\mu_{max},\mu_A^U/\mu_{max}]$. Note that the largest output intensity satisfies $\mu_A^U \leq \mu_{max}$, i.e. the transmittance is always no larger than 1.\\

Next, we make an observation here that, in scenarios where $\mu_A^U$ is strictly smaller than $\mu_{max}$, the local loss can be further divided into a constant part and a varying part. The constant part can be considered as a fixed beamsplitter BS1 with transmittance ${\mu_A^U \over \mu_{max}} < 1$, and the varying part can be considered as another beamsplitter BS2 with transmittance within $[\mu_A^L/\mu_A^U,1]$. An illustration of the two beamsplitters, along with the global phase shifter, is shown in Fig. \ref{fig:a7} (c)\footnote{Technically speaking, the phase shift $\phi_A$ is applied \textit{before} BS in Fig. \ref{fig:a7} (b), so to keep the equivalence, a phase shift $\phi_A$ should also be applied in Fig. \ref{fig:a7} (c) before BS1. Equivalently, two phase shifts of $\phi_A$ can be respectively applied after BS1 on the discarded port inside Alice's lab as well as before BS2. Note that the discarded port is considered part of the source and is kept within Alice's lab and away from Eve, so the phase shift to that port is actually irrelevant. The point is, there is no randomly varying phase shift to the signals exiting BS1, i.e. the ``source" is still an ideal source that outputs signals with fixed phases $0$ or $\pi$ like in active TF-QKD, just with a new intensity $\mu_A^U$ instead of $\mu_{max}$.}.

Here we emphasize the key physical insight: the entire setup in Fig. \ref{fig:a7} (a) or (b) is actually inside Alice's lab, i.e. trusted by Alice. She fully knows the description of the output state for each combination of phase slices (it's just that the states are not in the form of $\ket{\alpha}$ and $\ket{-\alpha}$ and have mixed intensities and phases). However, given the protocol description of CAL TF-QKD, it is difficult to directly apply its security analysis to these mixed source states. Therefore, we chose to take a more pessimistic but convenient approach and assume that Alice first sends perfect states, which pass through a local channel that is considered as part of the total channel loss and attributed to Eve.

However, we do not always have to attribute the \textit{entire} local channel to Eve. So long as we can keep a convenient pure-state description of the source (such that we can still apply the same security analysis from active CAL TF-QKD), it is up to us to decide how much Alice assumes her local channel to be untrusted and attributed to Eve, since the whole channel is hers to begin with.

As shown in Fig. \ref{fig:a7} (c), we notice that the state after BS1 is still a pure coherent state, so we can choose to only attribute the phase shift $\phi_A$ and BS2 to Eve, while keeping BS1 (and its discarded port) trusted and considered as part of Alice's source. Security-wise, this is a different and less pessimistic setup than Fig. \ref{fig:a7} (a) and (b) and should result in less privacy amplification, as it yields less information to Eve. Practically, this allows us to assume a smaller signal intensity for Alice and Bob, while maintaining the same physical observables in both X and Z bases, which can (1) reduce the cat state coefficients and give us a slightly better key rate even in the asymptotic case, and more importantly (2) allow weaker signal states than decoy states and thus enable finite-size analysis in Ref. \cite{finiteTF1} compatible with our protocol.

\subsection{Revised Post-Selection Strategy}

\begin{figure}[t]
	\includegraphics[scale=0.25]{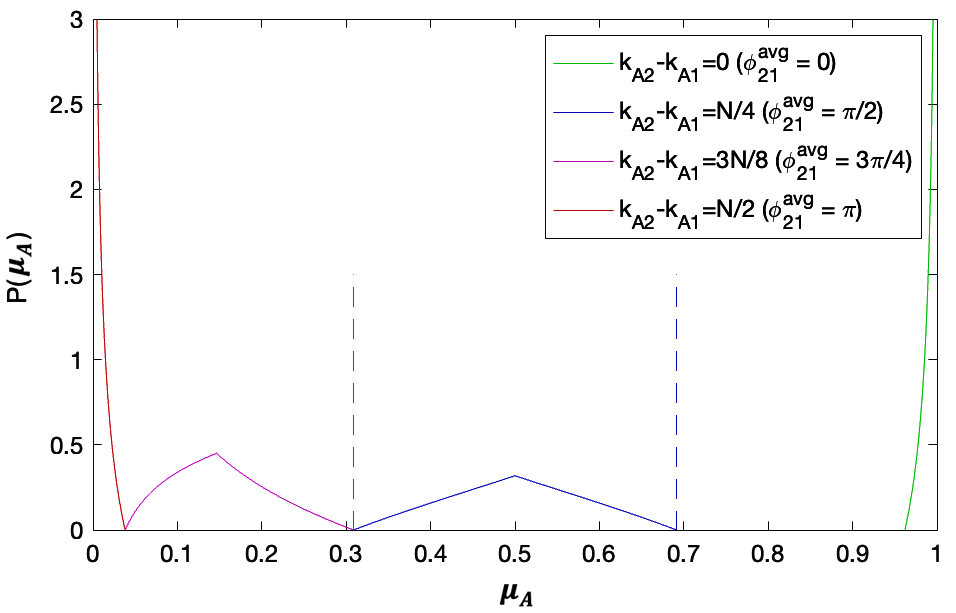}
	\caption{{\color{black}An illustration of the output intensity distribution $p(\mu_A)$ for different slice index combinations of $k_{A1}$ and $k_{A2}$. Here we take $N=16$ slices and $\mu_{max}=1$ as an example and show the output $[\mu_A^L,\mu_A^U]$ for four cases of $k_{A2}-k_{A1}\in \{0,N/4, 3N/8,N/2\}$,
		i.e. $\phi_{A2}^{avg}-\phi_{A1}^{avg}\in \{0,\pi/2,3\pi/4,\pi\}$. As can be seen, the intensity fluctuates within a quite large region within a pair of phase slices combination, and the maximum intensity $\mu_A^U$ is often quite large, say $\sim 0.3$ for slices $3\pi/4$ apart and $\sim 0.7$ for slices $\pi/2$ apart, meaning that using the revised model in Fig. \ref{fig:a7}(c) alone still cannot produce sufficiently small signal intensities. The non-linear shapes of $p(\mu_A)$ is due to $\phi_{21}=\phi_{A1}-\phi_{A2}$ having a triangular distribution and $\mu_A=\mu_{max}(1-cos\phi_{12})/2$ being a non-linear function. Also, $d\phi_{21}/d\mu_A$ is sometimes piece-wise, depending on whether the domain of $\phi_{21}$ within the phase slice combination covers $0$ or $\pi$, resulting in different shapes in $p(\mu_A)$.}}
	\label{fig:a11}
\end{figure} 

Having described an improved local channel model that decouples the local loss and the global phase, in this subsection we further describe a new post-selection strategy in the X basis that enables finer control of the signal intensities by directly post-selecting upon the local loss and the global phase independently. Note that here in the whole subsection we will focus on the X (key generation) basis, while the post-selection strategy for decoy states in the Z basis is identical to that of the asymptotic case, as described in the main text.

In the main text and Sec. D of the Appendix, we proposed a strategy where each key-generating pattern consists of two phase slices with indices $k_{A1},k_{A2}$ for Alice (and $k_{B1},k_{B2}$ for Bob), which are paired up into phase slice pairs indexed by $\{l_{A1},l_{A2},l_{B1},l_{B2}\}$ to generate key. However, while this strategy works well in the asymptotic case (which does not enforce any restriction on the signal intensity), in the finite-size case it does not have a fine enough control over the signal intensities. For instance, an illustration can be found in Fig. \ref{fig:a11}, where we take $N=16$ slices and $\mu_{max}=1$ as an example and plot the output $[\mu_A^L,\mu_A^U]$ when we combine signals that fall within two phase slices $k_{A1}$ and $k_{A2}$. For four cases of $k_{A2}-k_{A1}\in \{0,N/4, 3N/8,N/2\}$,
i.e. the difference in center positions of the two slices satisfies $\phi_{A2}^{avg}-\phi_{A1}^{avg}\in \{0,\pi/2,3\pi/4,\pi\}$, as we can see, except for phase slices identical or directly opposite to each other, for most phase slice combinations the output intensity fluctuates within a rather wide region, and $\mu_A^U$ is rather large, so even considering the improved channel model in Fig. \ref{fig:a7} (c) and combining the constant part of local loss to the virtual source, the signal intensity is still not small enough, and we need to find some better way to post-select smaller signal intensities.

Here we propose a new post-selection strategy for the key generation X basis that is different from that of the main text. We recall that, as we observed in Appendix F.2, before any post-selection is performed, Alice sends the same mixed states regardless of the X or Z bases. In this state, the global phase $\phi_G$ lies uniformly within $[0,2\pi)$, and the intensity $\mu_A$ follows a characteristic ``U-shaped" distribution $p_{int}(\mu)=1/(\pi \sqrt{\mu(\mu_{max}-\mu)})$. The phase and intensity are also decoupled and independent of each other. This means that, for the key generation X basis, in fact we can directly post-select the intensity within a region $[s^L,s^U]$ and further post-select the global phase within $[\phi_A^{avg}-\Delta, \phi_A^{avg}+\Delta]$. This is somewhat similar to what we are already doing in the decoy state Z basis, where we post-select regions in the intensity distribution but ignore the global phase. By explicitly controlling the intensity within $[s^L,s^U]$, we can limit the maximum intensity sent by Alice, and hence set a low value for $\mu_A^U$ (i.e. higher proportion of fixed local loss). Note that $\mu_A,\phi_A$ holds the same two degrees-of-freedom as $\phi_{A1},\phi_{A2}$, and in fact we can even in practice still measure $\phi_{A1},\phi_{A2}$ with the same physical setup but simply calculate $\mu_A,\phi_A$ before performing post-selection on them. The new asymptotic key rate becomes

\begin{equation}\label{eq:rate_new_strategy}
	\begin{aligned}
		R =&  \left(\int_{s_L}^{s_U} P_{int}(\mu) d\mu\right)^2 {4 \over {N^2}} \sum_{l_{A}=0}^{(N/2)-1} \sum_{l_{B}=0}^{(N/2)-1}\\
		&[\max(0,R_{0,1} (l_{A},l_{B})) \\
		&+ \max(0,R_{1,0} (l_{A},l_{B}))],\\
		=& \left(\int_{s_L}^{s_U} P_{int}(\mu) d\mu\right)^2 {2 \over {N}} \sum_{l_{B}-l_{A}=0}^{(N/2)-1}\\
		&[\max(0,R_{0,1} (0,l_{B}-l_{A}))\\
		&+ \max(0,R_{1,0} (0,l_{B}-l_{A}))],\\
	\end{aligned}
\end{equation}

\noindent where we have again used rotational symmetry (as described in Appendix D.2) to combine phase slice pair combinations $l_A,l_B$ that differ by only a global phase. Notably, with only two phase slices to post-select and with rotational symmetry applied, the number of unique combinations reduces to as few as $N/2$, which drastically reduces the impact of phase slice post-selection on the data size in the key-generating X basis.

One thing worth noting is that, in the asymptotic case, this strategy is not necessarily advantageous when compared to the phase slice matching strategy in the main text and in Appendix D, which is why we still used the latter for simulations in the main text for the asymptotic key rate. In the asymptotic scenario, having more phase slices always means better key rate. Directly post-selecting upon a single intensity region $[s^L,s^U]$ means that we are potentially grouping several phase slice combinations (that fall within the output intensity range) and calculating the gross key rate in one bin. This results in loss of fine-grained information and actually makes privacy amplification looser. However, for the finite-size case, post-selecting the intensity directly is more advantageous since it gives us better control over the output intensity and allows us to choose only small signal intensities, while, as shown in Fig. \ref{fig:a11}, the phase slice matching strategy results in widely fluctuating intensities in signal states. \footnote{Additionally, it is certainly possible to define multiple finer intensity regions and sum up the keys generated from each of them, but, again, such an approach will not be ideal for the finite-size scenario, while for the asymptotic scenario, having infinitely many intensity regions and having infinitely many phase slices are equivalent.}

\begin{figure}[t]
	\includegraphics[scale=0.38]{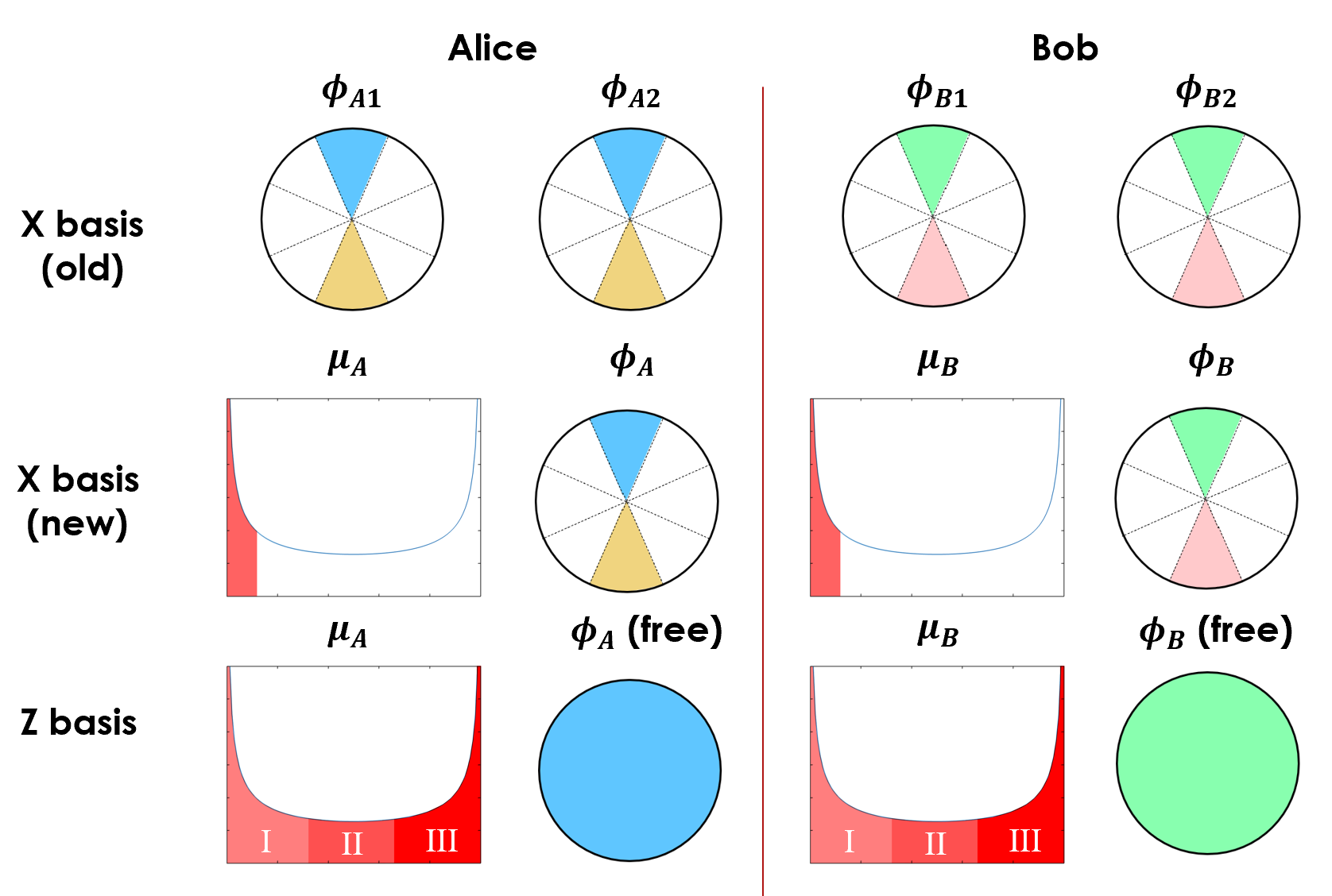}
	\caption{{\color{black}A comparison of the post-selection strategies in the main text asymptotic case (old) and here for the finite-size case (new). In the old X basis, Alice and Bob each post-select $\phi_{A1},\phi_{A2}$ and $\phi_{B1},\phi_{B2}$. In the new X basis, they instead post-select $\mu_A,\phi_A$ and $\mu_B,\phi_B$ (which contains the same degrees-of-freedom but allows us to choose smaller $\mu_A,\mu_B$ directly). Here for illustration the intensity post-selection region is chosen such that it starts from zero, but the lower boundary can be nonzero too. The Z basis strategy is identical, where Alice and Bob divide the entire intensity region into bins and post-select them, while keeping the phases uniformly random without post-selection.}}
	\label{fig:a12}
\end{figure}

\begin{comment}
	After Alice's post-selection in the X basis, for each key-generating phase slice pattern, $\phi_{A1}$ and $\phi_{A2}$ uniformly lie within two phase slices with center positions at $\phi_{A1}^{avg}=k_{A1}\times (2\Delta_{\phi})$ and $\phi_{A2}^{avg}=k_{A2} \times (2\Delta_{\phi})$ (where $k_{A1}.k_{A2}$ are the slice indices) and with boundaries of $[-\Delta_{\phi},\Delta_{\phi}]$ from the center positions. Here $2\Delta_{\phi}={{2\pi} \over N}$ is the width of each phase slice. 
	
	respectively taking random values within the intervals of $[\phi_{A1}^{avg}-\Delta_\phi, \phi_{A1}^{avg}+\Delta_\phi]$ and $[\phi_{A2}^{avg}-\Delta_\phi, \phi_{A2}^{avg}+\Delta_\phi]$ (corresponding to the phase slice choices).
\end{comment}

\subsection{Description of Finite-Size Analysis}

In this subsection, we largely recapitulate the results from the finite-size analysis from Ref. \cite{finiteTF1} and point out the modifications needed for the passive setup.

Firstly, the data flow to calculate the key rate in the asymptotic scenario is summarized as follows: Alice and Bob, with the help of the untrusted relay Charlie, first perform quantum communication and post-processing (including sifting, testing, and error-correction) to obtain the Gain in the X (key-generating) and Z (decoy) bases $Q^X,Q^Z_{ij}$ and the QBER in X basis $E^X$.  Note that here we follow the convention in Ref. \cite{finiteTF1} and combine the $(0,1)$ and $(1,0)$ events for $k_c,k_d$ detector clicks during parameter estimation and privacy amplification, in order to make sifted data blocks larger. The Z basis Gain data $Q^Z_{ij}$ is used to estimate upper bounds on the yields $Y^Z_{m_Am_B}$, which are further used to bound the X basis phase error rate $e^Z$. The upper bound for the key rate can be obtained from $Q^X,E^X$ and $e^Z$ and used in privacy amplification.

In the finite-size scenario, the data flow stays largely the same. However, one must factor in the statistical fluctuations in three steps: 

\begin{enumerate}
	\item The Z basis decoy state data $Q^Z_{ij}$ used to bound the yields $Y^Z_{m_Am_B}$ are not expected values but observed values, so one needs to bound the expected values of $Q^Z_{ij}$ within a confidence interval, using e.g. the Chernoff bound;
	\item The bit error rate in the Z basis $e^Z$ is not a perfect representation for the phase error rate in the X basis $e^X_{ph}$, given the finite size for both the X and Z basis data, so one must loosen the bound on $e^X_{ph}$ when calculating the key rate, e.g. in the random sampling case using Serfling's inequality and in the general case using Azuma's inequality or (as proposed in Ref. \cite{finiteTF1}) using a modified concentration inequality;
	\item Correction terms must also be added for the final key length, compared to the asymptotic key rate of $R=Q^X[1-h_2(e^Z)-h_2(E^X)]$, to ensure composable security.
\end{enumerate}

Here for reference we list some newly defined constant settings and intermediate variables that are used in the finite-size analysis:

\begin{itemize}
	\item The total data sent, $N_{s}$ (not to be confused with $N$, the number of phase slices);
	\item The basis choice probabilities, $p_X,p_Z$;
	\item The decoy state intensity choice probabilities, $p_i$;
	\item The received (and sifted) X and Z bases data, $M_X=N_sp_X^2Q^X$ and $M_Z=N_sp_Z^2 \sum_{ij}p_ip_jQ^Z_{ij}$;
	\item The total received data, $M_s=M_X+M_Z$;
	\item The conditional probability of sending $m_A,m_B$ photons given that Z or X basis is chosen, $p_{m_Am_B|Z}=\sum_{ij}p_ip_jP_{Poisson}(m_A,\mu_i)P_{Poisson}(m_B,\mu_j)$ and $p_{m_Am_B|X}=P_{Poisson}(m_A,s_A)P_{Poisson}(m_B,s_B)$ where $P_{Poisson}$ is the Poissonian distribution and $s_A,s_B$ are the signal intensities;
	\item The received data with $m_A,m_B$ photons in the Z basis, $M_{m_Am_B,Z}=N_sp_Z^2p_{m_Am_B|Z}Y_{m_Am_B}^Z$;
	\item The phase error correction terms, $\Delta_{m_Am_B}$;
	\item The security parameters (failure probabilities), $\varepsilon_c,\varepsilon_{a},\epsilon_{PA},\epsilon_{C}$, respectively for Chernoff bound, concentration inequality, privacy amplification in total, and correctness.
\end{itemize}

Now, we simply recapitulate the results for active TF-QKD from Ref. \cite{finiteTF1} below for points 1-3.

For point 1, one can bound the expectation values $\overline{Q}^{Z}_{ij}$ of the observables $Q^{Z}_{ij} $ using the Chernoff bound:

\begin{equation}\label{eq:Chernoff}
	\begin{aligned}
		N_{ij}^Z\overline{Q}^{Z,L}_{ij} &= {N_{ij}^Z Q^{Z}_{ij} \over{1+\delta^L}} \\
		&= -W_0\left(-exp\left({{ln(\varepsilon_C/2)-N_{ij}^Z Q^{Z}_{ij}}\over {N_{ij}^ZQ^{Z}_{ij}}}\right)\right)\\
		N_{ij}^Z\overline{Q}^{Z,U}_{ij} &= {N_{ij}^Z Q^{Z}_{ij} \over{1-\delta^U}} \\
		&= -W_{-1}\left(-exp\left({{ln(\varepsilon_C/2)-N_{ij}^ZQ^{Z}_{ij}}\over {N_{ij}^Z Q^{Z}_{ij}}}\right)\right)\\
	\end{aligned}
\end{equation}

\noindent with a failure probability of $\varepsilon_C$. Here $W_0,W_{-1}$ are Lambert W functions.

For point 2, the phase error rate is bounded by:

\begin{equation}
	\begin{aligned}
	e_{ph}M_X = N_{ph} \leq {p_X^2\over p_Z^2} \sum_{j=0,1} \left[y_{j} + y'_{j}\right]^2 + \Delta
	\end{aligned}	
\end{equation}

\noindent where two intermediate terms $y_j,y'_j$ are defined as:

\begin{equation}\label{eq:tail}
	\begin{aligned}
		y_{j} &= \sum_{\substack{m_A,m_B\in \mathbb{N}_j\\m_A,m_B\leq S_{cut}}} \sqrt{p_{m_Am_B|X} \over p_{m_Am_B|Z}} \sqrt{M_{m_Am_B,Z}+\Delta_{m_Am_B}} \\
		y'_{j} &= \sum_{\substack{m_A,m_B\in \mathbb{N}_j\\m_A,m_B > S_{cut}}} \sqrt{p_{m_Am_B|X}\over p_{m_Am_B|Z}} \sqrt{M_Z+\Delta}\\
		&= \sqrt{M_Z+\Delta} \sum_{\substack{m_A,m_B\in \mathbb{N}_j\\m_A,m_B> S_{cut}}} \sqrt{p_{m_Am_B|X}\over p_{m_Am_B|Z}}\\
	\end{aligned}	
\end{equation}

\noindent where $S_{cut}$ is a cutoff for the photon number, $\mathbb{N}_j$ represent pairs of odd-odd or even-even index pairs depending on $j$, and the correction terms are defined as

\begin{equation}
	\begin{aligned}
		\Delta & = \sqrt{{1\over 2} M_s ln{1\over \epsilon_A}} &\\
	\end{aligned}
\end{equation}

\noindent and

\begin{equation}
	\begin{aligned}
		\Delta_{m_Am_B} = \Delta \quad\quad\quad\quad\quad(m_A,m_B \neq 0)\\
		\Delta_{00} = \left[b+a\left({2M_{00,Z}\over M_s} - 1\right)\right]\sqrt{M_s} \;\;(m_A=m_B=0)
	\end{aligned}
\end{equation}

\noindent where $a,b$ can be arbitrarily chosen and are subject to optimization. One can also determine optimal values of $a,b$ analytically based on estimated knowledge of the channel and an informed guess of $M_{00,Z}$ (details can be found in Ref. \cite{finiteTF1}, Eq. 32).

For point 3, the key rate is corrected by:

\begin{equation}
	\begin{aligned}
		R\times N_s=l \leq &M_X [1-h_2(e_{ph}^U)] - M_X f h_2(e_X)\\
		&- log_2{2 \over \epsilon_C} - log_2{1\over {4\epsilon^2_{PA}}}
	\end{aligned}
\end{equation}

For the passive scenario, we can largely reuse the above analysis (given that the average intensity in the signal post-selection region is smaller than the largest of average intensities in the decoy setting regions), and only a few revisions need to be made, including that:

1. All observables will be obtained from integrating over a certain post-selection region, such as $\langle Q^X \rangle_{S_X}$, $\langle Q^XE^X \rangle_{S_X}$ and $\langle Q^Z_{ij} \rangle_{S_{ij}}$, where $\langle \rangle$ represents expected observable values over the integral region, $S_{k_Ak_B}$ represents an X basis key generation region (both Alice's and Bob's intensities are within $[s^L,s^U]$ and phase slice pairs $l_A,l_B$ are chosen) and $S_{ij}$ are decoy settings (defined by regions say I, II, III in Fig. \ref{fig:a12} for both Alice and Bob). The probability of choosing a decoy setting is $P_{S_{ij}}$, as defined in Eq. \ref{eq:decoy_prob}. The probability of choosing a signal setting, $P_{S_{k_Ak_B}}=({2\over N})\left(\int_{s^L}^{s^U} P_{int}(\mu) d\mu\right)^2$, will be discussed in point 3 below.

2. There is no basis sifting between Alice and Bob, since Alice directly chooses her bases randomly with probability $p_X,1-p_X$ and announces the results to Bob after transmission during post-processing, so $M_X\propto N_sp_X\langle Q^X \rangle_{S_X}$ (additional sifting due to post-selection will be discussed later) and $M_Z=N_sp_Z \sum_{ij}P_{S_{ij}} \langle Q^Z_{ij} \rangle_{S_i,S_j}$ for passive TF-QKD. The conditional probabilities of sending $m_A,m_B$ photons are also revised into $p_{m_Am_B|Z}=\sum_{ij}\langle P_{Poisson}(m_A,\mu_i)P_{Poisson}(m_B,\mu_j) \rangle_{S_{ij}}$ and $p_{m_Am_B|X}=\langle P_{Poisson}(m_A,s_A)P_{Poisson}(m_B,s_B)\rangle_{S_X}$.

3. There is, however, sifting due to post-selection. When Alice and Bob choose the X basis, each of them has a signal sifting probability of $P_{signal} = \int_{s^L}^{s^U} P_{int}(\mu) d\mu$, where the acceptable intensity region is $[0,s]$ and $p_{int}$ is the ``U-shaped" intensity probability distribution. Also, they perform privacy amplification and calculate key rates independently for each of their phase slice pattern $\{l_A,l_B\}$ based on $\phi_A,\phi_B$ and sum up the key rates. For each individual pattern, $M_X=N_sp_X({2\over N})P_{signal}^2 \langle Q^X \rangle$. On the other hand, they only need to perform decoy analysis once to obtain the upper bound on yields, so $M_Z$ is not affected, and the $Y_{n_An_B}$ obtained from finite-size analysis can be reused in all X basis phase slice patterns.

This is perhaps the biggest difference between passive and active TF-QKD in the finite-size regime, since, contrary to traditional TF-QKD or even BB84 where the key generation data size is generally equal to or larger than the testing/decoy-state data size, here for passive TF-QKD the X basis key generation data size is actually much smaller than that of the Z basis, potentially resulting in loosened bound on the phase error rate. However, as we will show through numerical simulation, even with the reduced X basis data size, at a reasonable phase slice number, satisfactory key rate (i.e. a tight enough bound on the phase error rate) can still be generated.

4. The local channels will result in corrections to the photon number yields, just like in the asymptotic case, and those corrections will naturally contain statistical fluctuations too. However, unlike for the external channel, we have perfect knowledge of the settings for those local channels inside Alice's and Bob's labs (hence the expected value of the transmittances of the local channels are known). In order to incorporate the local channels into the yields in our analysis, here we propose to take a slightly different approach from Ref. \cite{finiteTF1}:

Originally in Ref. \cite{finiteTF1}, the linear program for the decoy state analysis uses the observed counts in each intensity setting to bound the observed counts of $m_A,m_B$ photons, $M_{m_Am_B,Z}$, which are furthermore used to bound the observed phase error counts. However, here since we already \textit{know} the expected value of the local channel transmittances, it would be more convenient for us to get the expected value of the external channel, too. Here, we use a traditional decoy state analysis setup

\begin{equation}
	\begin{aligned}
		\sum_{m_A,m_B} P_{Poisson}(m_A,\mu_i) P_{Poisson}(m_B,\mu_j) \overline{Y_{m_Am_B}^Z} \geq \overline{Q}^{Z,L}_{ij}\\
		\sum_{m_A,m_B} P_{Poisson}(m_A,\mu_i) P_{Poisson}(m_B,\mu_j) \overline{Y_{m_Am_B}^Z} \leq \overline{Q}^{Z,U}_{ij}\\
	\end{aligned}
\end{equation}

\noindent to solve for the upper and lower bound of the expected value of yields, denoted as $\overline{Y_{m_Am_B}^Z}$ (in other words, expected values of $M_{m_Am_B,Z}$, instead of their observed values, divided by the total number of $m_A,m_B$ photons sent). Then, we can use Eq. \ref{eq:corrected_yield} to apply the expected values of transmittances of local channels to obtain the expected value of corrected yields, $\overline{Y^{'Z}_{n_An_B}}$. Once we obtain these expected values, we can apply the inverse of Eq. \ref{eq:Chernoff} (which is in the inverse multiplicative form and bounds expected values from observed values) and use it to bound the observed values $Y^{'Z}_{n_An_B}$ from the expected values. Then, we can simply use $Y^{'Z}_{n_An_B}$ instead in the phase error calculation (i.e. $M_{n_An_B,Z}$ will be replaced by $M'_{m_Am_B,Z}=N_sp_Zp_{m_Am_B|Z}Y^{'Z}_{m_Am_B}$ that represents the overall corrected yields) as they are already the physically observed counts that contain fluctuations.

Again, the above is only a simple proof-of-principle analysis aimed at estimating the performance of our passive TF-QKD protocol under finite-size effects, and it is more of a ``patch" (that first estimates and manipulates expected values and finally converting them to bounds on observed values) to make the additional local channel model compatible with the analysis in Ref. \cite{finiteTF1}. There could potentially be more elegant solutions (such as bounding the fluctuations of the local channels directly and combining the result with observed $M_{m_Am_B,Z}$) as well as a more rigorous study, which will be subject of future work.

\subsection{Simulation Results}

Here we perform a numerical simulation for passive TF-QKD under finite-size effects, as shown in Fig. \ref{fig:a13}. Dark count probability is set to $10^{-8}$, error-correction efficiency is set to $f=1$, and detector efficiency is incorporated into the channel loss (which assumes $\alpha=0.2$dB/km). Slice number is set to $N=16$. We choose a constant $\varepsilon=10^{-7}$ and a factor of $C_\varepsilon = (S_{cut}/2+1)^2 + 1 + N_{decoy}$, where the number of decoy settings is $N_{decoy}=3\times 3$ and photon number cutoff is set to a large $S_{cut}=30$ (such that the tail terms in $y_j'$ is effectively negligible, as $p_{m_Am_B|X} / p_{m_Am_B|Z}$ is close to or smaller than $10^{-14}$ beyond the cutoff), and we set the security parameters to $\epsilon_{C}=\varepsilon$, $\epsilon_{PA}=\varepsilon/3$, and $\varepsilon_a = \varepsilon_c = (\varepsilon/3) \times (1/C_\varepsilon)$. The active TF-QKD parameters including signal intensity $s_A,s_B$ , decoy intensities $\mu_i,\mu_j$ and their probabilities $p_i,p_j$, basis choice probability $p_X$, as well as the passive TF-QKD parameters including maximum signal intensity $\mu_{max}$, signal regions $[s^L,s^U]$, decoy regions $S_{ij}$, basis choice probability $p_X$, are all subject to optimization.

As can be seen, even in the finite-size scenario and with reasonable data size, passive TF-QKD is still able to yield a considerable amount of key rate despite the heavy post-selection performed, e.g. the small phase slices. This is thanks to two main factors (1) the new strategy in Fig. \ref{fig:a12} that only matches two local phases slices, instead of four (2) the utilization of symmetry that groups phase slice combinations that only differ by a global phase.

\begin{figure}[t]
	\includegraphics[scale=0.20]{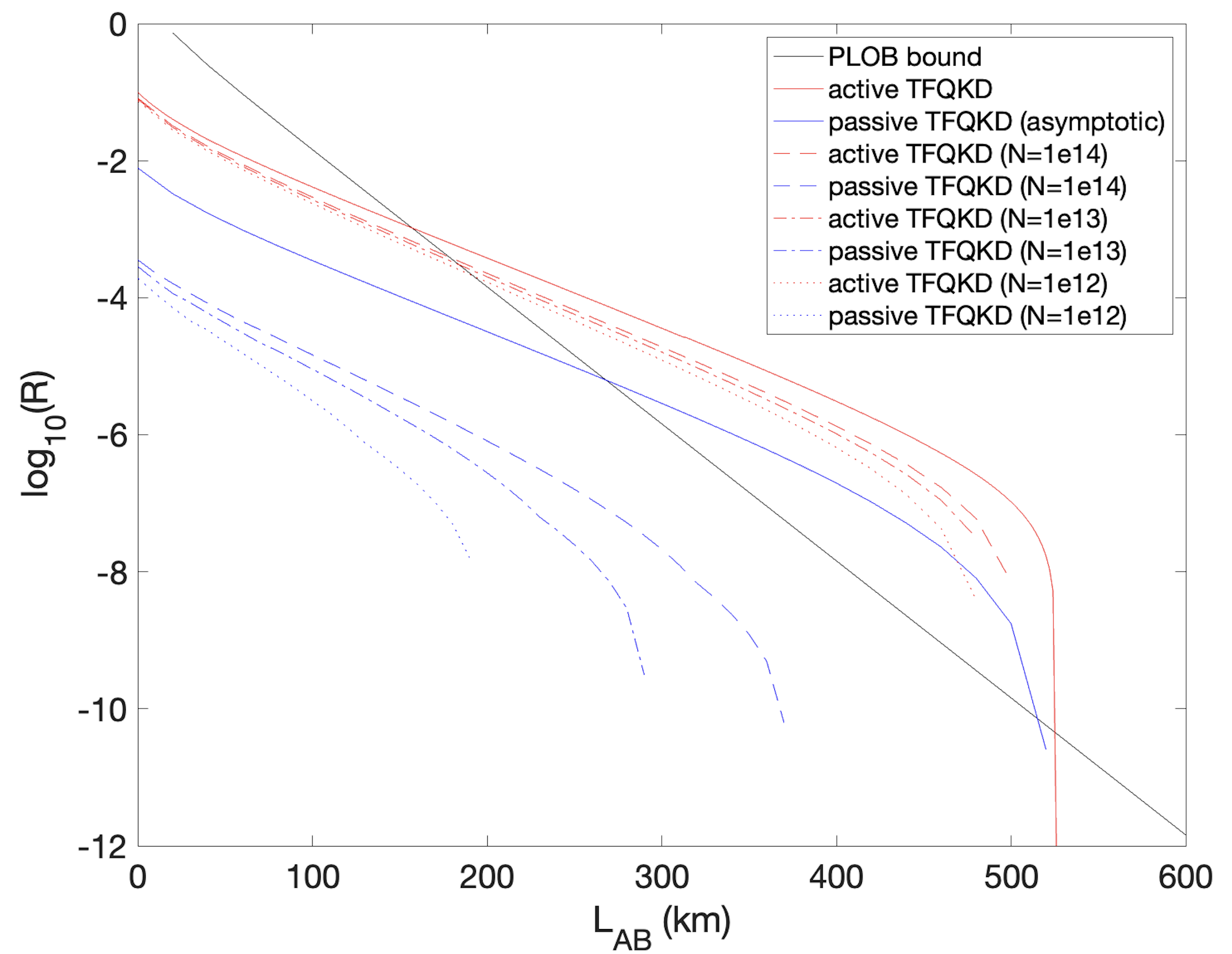}
	\caption{{\color{black}A comparison of the key rate for active and passive TF-QKD in asymptotic and finite-size scenarios. The asymptotic data is generated in the same way as that in the main text (using the phase slice matching strategy and assuming infinite number of decoys, i.e. perfect knowledge of yields). To obtain the finite-size data, for active TF-QKD the users choose from three decoy settings in the Z basis in active TF-QKD, and for passive TF-QKD the users employ the new post-selection strategy as shown in Fig. \ref{fig:a12} in the X basis.}}
	\label{fig:a13}
\end{figure}

Nonetheless, it does seem that the gap between asymptotic and finite-size cases for passive TF-QKD is quite a bit larger than for active TF-QKD (which suffers very small loss from finite-size effects). This is mainly because the signal intensity in the passive case, even with the new security boundary model in Fig. \ref{fig:a7} reducing it, is still on the same order of magnitude as the highest decoy intensity. This causes the tail terms in Eq. \ref{eq:tail} (higher photon number contributions to the phase error rate) to converge much more slowly in the passive case compared to the active case. Another less significant reason is that, unlike most active QKD schemes, in the passive case, the number of X basis signals in each key-generating bin is actually smaller (by one order-of-magnitude) than the Z basis. This results in a further hit on the key rate when applying concentration/Azuma's inequality to bridge X and Z data (which depends on both $M_X$ and $M_Z$). This analysis, however, mainly serves as a proof-of-principle that our new passive TF-QKD scheme can work even under the finite-size regime. We expect better key rate from finer optimization of the parameters, tailored revision to the bounds (rather than the current bounds we use that is designed mainly for active TF-QKD where $s >> \mu$), or testing a different proof framework (such as Refs \cite{finiteTF2,finiteTF3}). Such optimizations, along with a rigorous, full finite-size analysis, will be the subject of future work.

}

\newpage


\begin{thebibliography}{}
\bibitem{footnote2} {\color{black} Using Alice as an example, $\{l_{A1},l_{A2}\}$  is actually different from $\{l_{A1},l_{A2}+N/2\}$, e.g. when $N=8$, Alice encoding with slice pairs $\{0,0\}$ and $\{4,4\}$ is different from using pairs $\{0,4\}$ and $\{4,0\}$. On the other hand, $\{l_{A1}+N/2,l_{A2}+N/2\}$ is the same as $\{l_{A1},l_{A2}\}$. Overall, there are $N^4/4$ unique sets of $\{l_{A1},l_{A2},l_{B1},l_{B2}\}$.}
\bibitem{footnote6} For two continuous-wave lasers, locking them to the same frequency naturally means that the phase difference between them is fixed. However, gain-switched lasers randomly initialize the phases every time they turn on. Importantly, having close to identical frequencies at the stable regions of each pulse (i.e. excluding the rising region and chirping) in principle does not contradict the statement that the pulses themselves (between either neighboring pulses across time, or remote pulses across space between Alice and Bob) can have independent and random phases. Therefore, the frequency stabilization is not a fundamental limitation, but rather a practical implementation/engineering challenge.
\end{thebibliography}
\end{document}